\documentclass[fleqn,usenatbib]{mnras}

\usepackage{newtxtext,newtxmath}
\usepackage[T1]{fontenc}

\DeclareRobustCommand{\VAN}[3]{#2}
\let\VANthebibliography\thebibliography
\def\thebibliography{\DeclareRobustCommand{\VAN}[3]{##3}\VANthebibliography}

\usepackage{graphicx}	% Including figure files
\usepackage{amsmath}	% Advanced maths commands
\usepackage{booktabs}
\usepackage{tablefootnote}
\usepackage{soul}
\title[Arrival Times Spectro-Temporal Analysis]{Spectro-temporal analysis of ultra-fast radio bursts using per-channel arrival times}

\author[M. A. Chamma et al.]{
Mohammed A. Chamma,$^{1,2}$\thanks{E-mail: mchamma@uwo.ca}
Victor Pop$^{2}$ and
Fereshteh Rajabi$^{2}$
\\
$^{1}$Department of Physics and Astronomy, The University of Western Ontario, 1151 Richmond Street, London, Ontario N6A 3K7, Canada\\
$^{2}$Department of Physics and Astronomy, McMaster University, Hamilton, Ontario L8S 4L8, Canada
}

\date{Accepted XXX. Received YYY; in original form ZZZ}

\pubyear{\the\year{}}

\begin{document}
\label{firstpage}
\pagerange{\pageref{firstpage}--\pageref{lastpage}}
\maketitle

\begin{abstract}
Fast radio bursts (FRBs), especially those from repeating sources, exhibit a rich variety of morphologies in their dynamic spectra (or waterfalls). Characterizing these morphologies and spectro-temporal properties is a key strategy in investigating the underlying unknown emission mechanism of FRBs. This type of analysis has been typically accomplished using two-dimensional Gaussian techniques and the autocorrelation function (ACF) of the waterfall. These techniques are effective and precise at all duration scales, but can be limited in the presence of scattered tails, complex morphologies, or recently observed microshot forests. Here, we present a technique that involves the tagging of per-channel arrival times of an FRB to perform spectro-temporal measurements using a Gaussian profile model for each channel. While scattering and dispersion remain important and often dominating sources of uncertainty in measurements, this technique provides an adaptable and firm foundation for obtaining spectro-temporal properties from all types of FRB morphologies. We present measurements using this technique of several hundred bursts across 12 repeating sources, including over 400 bursts from the repeating sources FRB 20121102A, FRB 20220912A, and FRB 20200120E, all of which exhibit recently observed microsecond-long ultra-FRBs, as well as 143 multi-component drift rates. In addition to retrieving the known relationship between sub-burst slope and duration, we explore other correlations between burst properties. We find that the sub-burst slope law extends smoothly to ultra-FRBs, and that ultra-FRBs appear to form a distinct population in the duration-frequency relation.
\end{abstract}


%%%%%%%%%%%%%%%%%%%%%%%%%%%%%%%%%%%%%%%%%%%%%%%%%%

%%%%%%%%%%%%%%%%% BODY OF PAPER %%%%%%%%%%%%%%%%%%

\section{Introduction}

Fast radio bursts (FRBs) exhibit a diverse and rich set of information through their spectro-temporal properties, often yielding clues to the underlying physical processes of these extremely distant, energetic, and short-lived (ranging from nanoseconds to seconds) coherent bursts. 

FRBs are broadly categorized into repeating and non-repeating sources, which exhibit different statistical properties in their burst bandwidths and durations \citep{Pleunis2021a,Petroff2022}. This observation precludes the possibility that some non-repeating sources have simply not yet been observed to repeat.

Repeating sources offer opportunities for continued monitoring and source localization, and often emit bursts with complex time-frequency structure not displayed by non-repeating sources. Examples of this structure include the presence of multiple sub-components within a single burst event, as well as the observed tendency for later components to drift to lower frequencies (the ``sad-trombone effect''; \citealt{Hessels2019}). This effect has been analogously observed within a single burst, where lower frequency components arrive slightly delayed relative to higher frequencies, referred to as the sub-burst slope law and/or intra-burst drift rate  \citep{Rajabi2020,Chamma2021,Chamma2023,Jahns2023,Brown2024}. This effect is observed even after the frequency dependent ($\propto \nu^{-2}$) delaying influence due to interstellar dispersion is removed, although the dispersion measure (DM) used can greatly affect measurements when characterizing drift rates and sub-burst slopes. 

Spectro-temporal properties, such as the duration, bandwidth, frequency, energy, drift rate and sub-burst slope, reveal interesting and unexpected relationships between one another that serve as constraints on physical emission or source models. For example, \citet{Hewitt2022} observed two distinct groups of bursts in the width-bandwidth-energy parameter space for bursts from repeater FRB 20121102A \citep{Spitler2016,Scholz2016,Marcote2017,Bassa2017,Chatterjee2017,Tendulkar2017}. Multiple studies have now observed a bimodal distribution in burst wait times \citep[e.g.][]{Li2021,Hewitt2022,Jahns2023,Zhang2023,Konijn2024}, as well as in burst energies \citep{Li2021} for FRB 20121102A.

A particularly strong correlation, which is a primary focus of this study, has been observed between the sub-burst duration and sub-burst slope/intra-burst drift in bursts from multiple repeating sources. First seen strongly in bursts from FRB 20121102A \citep{Rajabi2020,Chamma2023,Jahns2023}, it appears that multiple other repeating sources exhibit the same relationship and even share similar scalings, namely FRB 20121102A, FRB 20180916B, FRB 20180814A. Others, like FRB 20201124A, show the same trend but may have different scalings, and more analysis is required \citep{Chamma2021,Wang2022,Brown2024}. This correlation is predicted by the triggered relativistic dynamical model (TRDM) whereby the FRB source, moving at up-to relativistic velocities relative to an observer, is triggered by an energy source (such as emission from a magnetar) located behind it and along the line of sight \citep{Rajabi2020}. In other words, the relationship can be interpreted as the direct result of dynamical (relativistic) motions in the FRB source that modulate the signal through the Doppler effect and differing reference frames between the source and the observer. This relationship takes the form of $\text{d}t/\text{d}\nu \propto \sigma_t$ where $\text{d}t/\text{d}\nu$ is the inverse sub-burst slope, or the change in burst arrival time with frequency, and $\sigma_t$ is the sub-burst duration. In earlier works \citep{Rajabi2020,Chamma2021,Chamma2023}, $\text{d}\nu/\text{d}t$ was used as the sub-burst slope, however, this leads to a singularity for bursts with little to no drift in time. Therefore, we adopt the recommendation of \citet{Jahns2023} to formulate the measurement as the inverse instead. Hereafter we will interchangeably refer to $\text{d}t/\text{d}\nu$ as the sub-burst slope or intra-burst drift.

As the number of bursts and the number of repeaters analysed grows, the robustness of the sub-burst slope relation ($\text{d}t/\text{d}\nu \propto \sigma_t$) is increasingly demonstrated. In the course of such analyses, there is growing evidence from the results of \citet{Chamma2023}, \citet{Jahns2023}, and \citet{Brown2024} that an analogous and/or identical relation exists for the drift rates of multiple components of an FRB and their overall duration. The interpretation of this is not clear but suggests a connection between the physical process that gives rise to drift within sub-bursts and the process that causes drift among distinct burst components.

Recently, multiple studies have reported the discovery of ultra FRBs --FRBs of microsecond or even nanosecond durations-- in bursts from FRB 20121102A \citep{Snelders2023} and FRB 20200120E \citep{Nimmo2022,Nimmo2023,Pearlman2024}. Additionally, bursts featuring dense forests of microshots have been observed from FRB 20220912A, a recently discovered and highly active repeater \citep{Hewitt2023,Zhang2023,Sheikh2024}. Understanding where the spectro-temporal properties of these ultra-FRBs fall is a valuable test of known spectro-temporal relationships and extends the measured parameter space by several orders of magnitude. Deviations (or the lack thereof) that may be observed are also significant, as these three sources are localized to disparate environments. For example, FRB 20200120E is localized to a globular cluster \citep{Kirsten2022}, while FRB 20220912A lacks a persistent radio source  \citep{Hewitt2024} seen in other repeating sources such as FRB 20121102A.

Gaussian formalisms have been used thus far in the literature to perform spectro-temporal analyses of bursts, including measurements of drift rate and sub-burst slope \citep[e.g.][]{Hessels2019,Rajabi2020,Chamma2021,Chamma2023,Jahns2023,Brown2024}. In \citet{Hessels2019}, a two-dimensional (2D) Gaussian model is fitted to the 2D autocorrelation function (ACF) of the burst waterfall, and burst properties are calculated from the model parameters. \citet{Jahns2023} introduced several 2D Gaussian models that could be fitted directly to the waterfall of an FRB and used parameters with physically consistent dimensionality; i.e., the burst properties are parameters of the Gaussian model, and the fits directly yield a measurement. Other modelling techniques that produce accurate spectral models of FRBs exist, such as \texttt{burstfit} \citep{Aggarwal2021} and \texttt{fitburst} \citep{Fonseca2023}; however, these do not lend themselves easily to acquiring sub-burst slope measurements due to their underlying formalism. Finally, while the Gaussian methods are effective at any duration scale, the models can be limiting if one wishes to study the spectro-temporal properties of bursts with non-gaussian, frequency-varying profiles or with complex morphologies and blended components.

This work will present a technique that relies on tagging the arrival time of an FRB signal in each frequency channel of its waterfall, using criteria based on a profile model of the channel. Using the arrival time data, one can apply various models to characterize the drifting behaviour of a burst. We apply a simple linear model to define the sub-burst slope measurement. This method is detailed in Section \ref{sec:methods}. We apply this method to obtain measurements for hundreds of FRBs from 12 different repeating FRB sources, with a strong focus on the three aforementioned repeaters that have exhibited ultra-FRBs or microshot behaviour: FRB 20121102A \citep{Snelders2023}, FRB 20220912A \citep{Hewitt2023,Zhang2023}, and FRB 20200120E \citep{Nimmo2022,Nimmo2023,Pearlman2024}. The data used are described and summarized in Section \ref{sec:data}. Section \ref{sec:results} details the results of these measurements, including figures showing the sub-burst slope relation and other correlations, as well as presenting drift rate measurements from bursts with multiple components. Section \ref{sec:results} also includes a comparison between the arrival times and Gaussian methods and highlights factors that can lead to differing measurements. 

Finally, in Section \ref{sec:discussion}, we discuss the interpretation of our results, possible extensions of the arrival times method, and directions for future analyses. The paper is summarized in Section \ref{sec:conclusion}. An abridged table of burst measurements can be found in Appendix \ref{app:ultrafrbs}, sorted by shortest duration. Appendix \ref{app:microshot} provides an example measurement of a microshot forest.

\section{Methodology}\label{sec:methods}

\subsection{Arrival times pipeline}\label{subsec:arrtime}

The arrival times pipeline finds the arrival time of an FRB in each frequency channel of its waterfall. It then uses these times to perform sub-burst slope measurements. Filtering for arrival times in channels with sufficiently high S/N (we chose 3$\sigma$) helps with measurement accuracy and quality.

The outline of the procedure is as follows:

\begin{enumerate}
\item Given a waterfall, first compute the 1D time series integrated over all frequency channels.
\item Fit a 1D Gaussian to the time series. Burst components are counted visually from the waterfall and time series. If multiple components (sub-bursts) are present in the waterfall, model the time series as the sum of multiple 1D Gaussians. The width, $\sigma_t$, of this 1D Gaussian is used as the duration measurement of the burst (or sub-burst component if there are multiple components).
\item \label{item:cutout}For each component of the 1D time series, use a multiple of its width (we chose $4\sigma_t$) to select a time range on either side and cut out the burst component from its waterfall.
\item For each frequency channel (row) of the waterfall, fit a 1D Gaussian to model the pulse in that frequency channel. The 1D Gaussian model used throughout (including for the time series fit) is given by:
\begin{equation} % or \begin{align}
    G(t)=A\exp\Big(-\frac{(t-t_{0})^{2}}{2\sigma^{2}}\Big),\label{eq:profile}
\end{equation}
with parameters $A$, $t_{0}$, and $\sigma$. The ``arrival time'' of the FRB in that frequency channel is then defined as 
\begin{equation}
    t_{\text{arr}} \equiv t_{0}-\sqrt{2}\sigma,\label{eq:arrtime}
\end{equation}
which corresponds to the $e^{-1}$ amplitude of the Gaussian. This point in time is found for each frequency channel.
% for me when rewriting: some outliers were present due to overaggressive downsampling 
\item \label{item:sn}Apply spectral and temporal filters to arrival time data: we find the S/N of the pulse in each frequency channel using the mean of pixels in the on-pulse region and the standard deviation of any pixels in an off-pulse region in the same channel, while ensuring that the noise is sampled over the same number of pixels used to measure the signal. If the S/N is greater than 3$\sigma$ in that channel, the arrival time in that channel is considered acceptable and retained for use in subsequent steps. The on-pulse region for each component is determined using the position $t_{0,\text{1D}}$ and width $\sigma_{t,1\text{D}}$ from the 1D Gaussian fit obtained from the integrated time series. This is generally sufficient for ensuring that arrival times accurately correspond to their burst in the waterfall. Nonetheless, an additional filter is applied to remove arrival times at durations greater than 2$\sigma_{t,1\text{d}}$ away from the component time obtained from the integrated time series fit. This last filter can especially help with blended components.

\item Using the arrival times that remain after applying the filters, a linear fit of the form 
\begin{equation}
    t_{\text{arr}}=\frac{\text{d}t}{\text{d}\nu}\nu + t_b\label{eq:arrtimefit}
\end{equation}
is performed. This defines the sub-burst slope $\text{d}t/\text{d}\nu$\footnote{This is identical to the measurement of $(\text{d}\nu_{\text{obs}}/\text{d}t_\text{D})^{-1}$ and $d_t$ from the formalisms of \citet{Rajabi2020} and \citet{Jahns2023}, respectively.} and is how the measurement is obtained. The intercept value $t_b$ is a constant offset necessary for a general linear model to accurately converge. Technically, $t_b$ represents the `arrival time' relative to the peak of the pulse if the burst extended to a frequency of 0, which, of course, is not observed. To further illustrate this, the parameter $t_b$ can be removed if $\nu$ is replaced with $\nu-\nu_\text{offset}$, where $\nu_\text{offset}$ is a different offset parameter, now in units of frequency, that is also generally numerical. Both choices result in the same measured value of $\text{d}t/\text{d}\nu$.
\item The burst duration $\sigma_{t}$ and bandwidth $\sigma_{\nu}$ of each burst are obtained from the 1D Gaussian fits of the integrated time-series and spectrum of each component, respectively. The spectrum is normally found by integrating over the 1$\sigma_{t\text{,1D}}$ width of the on-pulse region, but can be configured to larger widths if, for example, the burst is well resolved and no blending is present. The center frequency $\nu_0$ of the FRB is defined as the peak of the fit to the integrated spectrum. In some cases, intensity variations due to scintillation can interfere with the measurement, resulting in an underestimated value for the burst bandwidth. For example, a burst may have a high intensity over only a small fraction of its total bandwidth. In such cases, some channels are masked based on a manually chosen intensity threshold and ignored during fitting, which helps yield a broader, more accurate bandwidth. This threshold is chosen by visual inspection of the spectrum. For example, in one burst, some scintles occupying a small portion of the spectrum may reach a peak intensity of 0.4 arbitrary units, while the majority of the signal is below 0.1 units. In this case, we may choose a threshold of 0.11 to indicate that frequency channels with intensities above this threshold, containing the scintles, are ignored during fitting.
\end{enumerate}

The burst duration $\sigma_t$ measured from the integrated time-series can be artificially broadened by drifting. We can alternatively define the burst duration as the average of the $\sigma$'s found in each frequency channel with sufficient S/N (as defined in step \ref{item:sn}). However, this may be limited by the S/N and the morphology of the burst. We also discuss using the measured slope to `correct' the duration by removing residual drift in Appendix \ref{subsec:dmopt}.

An additional step is performed in the presence of multiple burst components to measure the drift rate between components. Using the center frequency $\nu_0$ and the time of each component, obtained from the integrated 1D spectrum and time series, a line analogous to that in Equation \ref{eq:arrtimefit} is fitted to define the drift rate $\Delta t/\Delta \nu$. We measure the `total duration' of an FRB with multiple components, which may have unpredictable spacings between them, as the time difference between the first and last components. This estimate of the duration turns out to be sufficient for our purposes, though other definitions may be required depending on the research goals.

For each fit performed, including the profile fit in each frequency channel, the 1D time series fit, the 1D spectrum fit, the sub-burst slope fit, and the drift rate fit, the residuals are computed using an appropriately sampled data uncertainty and used to compute the reduced-$\chi^2$ ($\chi^2_{\text{red}}$). Fit parameters are obtained from the non-linear least squares routine implemented by the \texttt{scipy.optimize.curve\_fit} package. By default, the covariance matrix provided by \texttt{curve\_fit} is based on scaling the given data uncertainty such that $\chi^2_{\text{red}}=1$ after the fit. In the measurements to come, we report the 1$\sigma$ uncertainties derived from this scaled covariance matrix alongside the $\chi^2_{\text{red}}$.

\begin{figure}
\textbf{(a)}

\hspace{-0.5cm}\includegraphics[width=1.1\columnwidth]{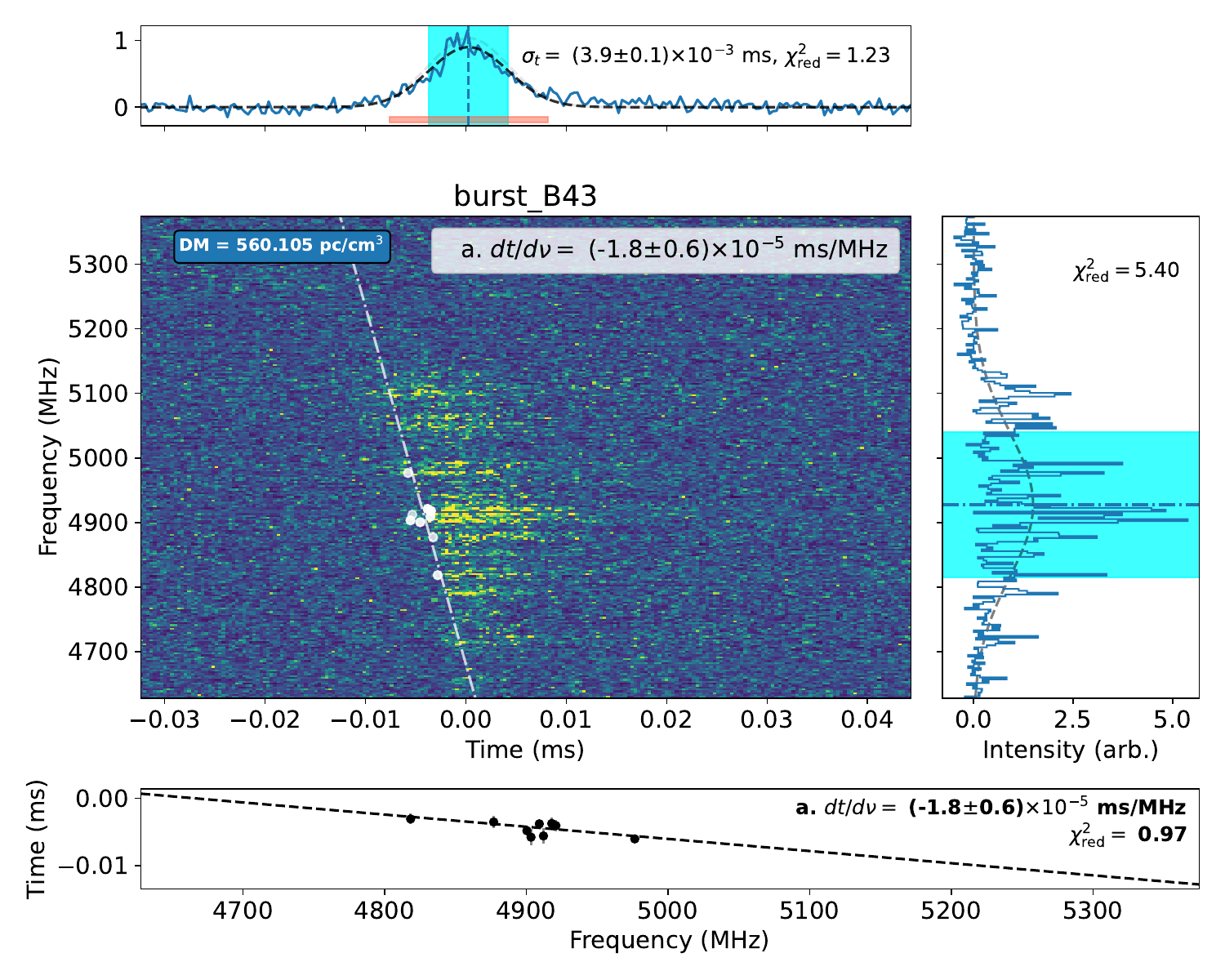}
\textbf{(b)}

\hspace*{-0.5cm}\includegraphics[width=1.1\columnwidth]{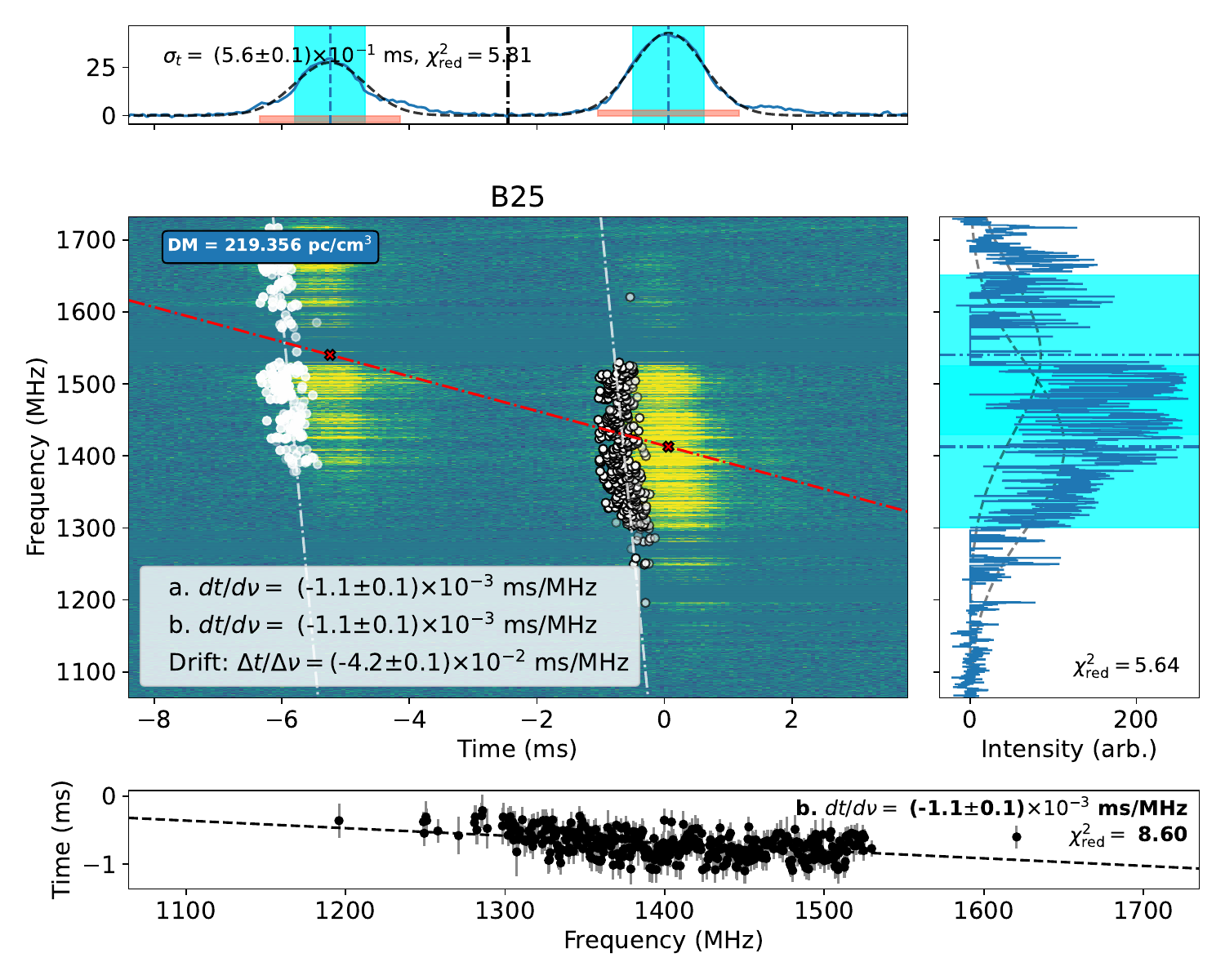}
\caption{\label{fig:arrmethod}\textbf{(a)} Arrival times method for burst B43 of \citet{Snelders2023}.  The top and rightmost panels show the integrated time series and spectrum, respectively, of the waterfall, shown in the center. In the time series, the 1D Gaussian fits are shown with a black line. The vertical dashed line denotes the peak time of the burst. The reddish bar indicates the 2$\sigma_{t,\text{1D}}$ window used in the arrival times temporal filter. The faint blue shaded regions in the time series and spectrum are the 1$\sigma$ regions of their corresponding fits. The spectrum for this burst is found by integrating over the 2$\sigma_{t,\text{1D}}$ region of the pulse. The dash-dot line denotes the center frequency. The waterfall is displayed with white points indicating the arrival times found in each frequency channel, which are then used in the linear fit to obtain $\text{d}t/\text{d}\nu$, shown in the bottom sub-panel. For each fit, the corresponding reduced-$\chi^2$ is shown in its corresponding panel. \textbf{(b)} Same as above but for burst B25 of \citet{Sheikh2024,Sheikh2024a}, which has two sub-bursts. The plot shows how different components are separated for measurement. Components are labeled alphabetically from left to right and each set of arrival time points are colored according to the sub-burst they are associated with. The red line indicates the drift rate measurement $\Delta t / \Delta \nu$ obtained for this burst and the red xs denote the center frequency of each component. The bottom sub-panel shows the linear fit and arrival times of the last component.}
\end{figure}

Figure \ref{fig:arrmethod} illustrates the method applied to both an FRB with a single component and an FRB with multiple components, including a drift rate measurement. Appendix \ref{app:ultrafrbs} provides an abridged table of measurements sorted by shortest duration.

\begin{figure}
\hspace{-0.5cm}\includegraphics[width=1.1\columnwidth]{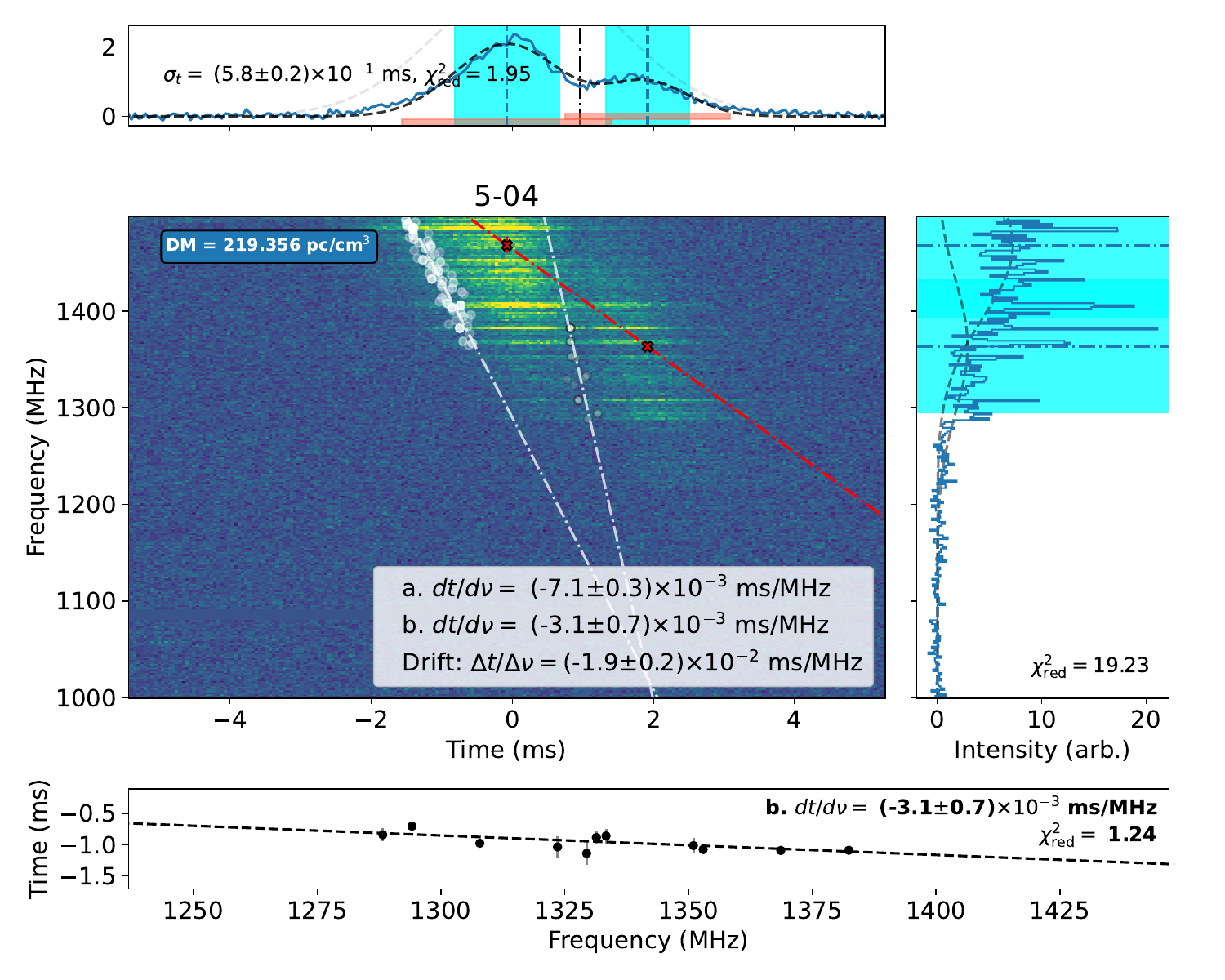}
\caption{\label{fig:arrmethod2}Burst 5-04 from \citet{Zhang2023} from FRB 20220912A. Same as Figure \ref{fig:arrmethod}, but showcasing measurements from two sub-components blended together. The black dash-dot line in the time series indicates the position of the manual cut. The drift rate shown includes a component truncated by the observing band, and will be marked specially in figures to come. The bottom sub-panel shows the linear fit and arrival times of the last component.}
\end{figure}

There are many instances of bursts with components that overlap or blend together. In such cases, we can still obtain reasonably good measurements by manually selecting where to cut out the bursts, rather than using the 4$\sigma_t$ width used in step \ref{item:cutout}. By using multiple manual cuts, we can separate many components and obtain measurements. This strategy can even be applied to microshot forests from, for example, \citet{Hewitt2023}, where one burst has over 40 components, with significant success. However, this process is still limited by the amount of blending in the waterfall and the ability to visually distinguish components. Figure \ref{fig:arrmethod2} shows an example of measurements from a blended two-component burst. Appendix \ref{app:microshot} provides an example measurement and additional details on the process and challenges of analysing heavily blended microshot waterfalls using the arrival times method.

The arrival times method provides certain advantages. In particular, the method naturally allows for different models of the relationship between $t_{\text{arr}}$ and frequency to be investigated. This is particularly relevant in light of the complex burst morphologies observed by \citet{Faber2023}, who argue for power law models to describe the drifting morphologies they observe. 

The arrival times pipeline is implemented in Python and packaged with \textsc{Frbgui}\footnote{\url{https://github.com/mef51/frbgui}} (though currently only in script form) and is also accessible via the Zenodo link provided at the end of the paper. Technical documentation is also available\footnote{\href{https://frbgui.readthedocs.io/en/latest/documentation/doc-arrivaltimes.html}{\texttt{https://frbgui.readthedocs.io/arrivaltimes}}}.

Note that this method, along with the other methods mentioned here, does not consider the covariance of the measurements with the dispersion measure (DM) or with scattering effects. Typically, a strategy must be applied before performing measurements to account for these significant and often dominant sources of uncertainty. In the case of the DM, this has typically been done by either: (a) performing measurements at each burst's independently obtained DM, as is done in a majority of the literature; (b) repeating measurements over broad ranges of possible DM values, effectively trading precision for accuracy (e.g., \citealt{Chamma2021,Chamma2023,Brown2024}); or (c) assuming the validity of the predicted relationship between $t_{\text{arr}}$ and $\nu$ (i.e., the sub-burst slope law of \citealt{Rajabi2020}) to correct the DM such that the deviation from this relationship is minimal (see \citealt{Jahns2023} for a quantitative application of this, and \citealt{Chamma2023,Brown2024} for a qualitative application). We will explore here another strategy: using the highly precise DMs of the newly discovered high frequency ultra-FRBs to select and impose the DM on the remaining bursts in a sample. We will find that, while this is effective in retrieving expected spectro-temporal relationships, it does not provide any signficant advantages over other methods, and is sensitive to the effects of interstellar scattering and dispersion. Other possible strategies could involve using the distribution of DMs from a source to better quantitatively estimate the covariance with spectro-temporal properties.

\subsection{Measurement filtering}\label{subsec:filters}

In order to accurately quantify the relationships between the spectro-temporal measurements obtained from the arrival times pipeline, we apply additional filters to the resulting measurements to exclude those with large uncertainties or invalid values. These typically include measurements from faint, blended, affected by radio frequency interference (RFI), or otherwise obscured burst pulses.

The primary filters applied remove measurements of bursts with relative uncertainties on their duration or sub-burst slope larger than 100\%. While these measurements can still be physically meaningful (e.g., a measurement of a burst with a slope of $2\times10^{-4}$ ms/MHz with $\sim$100\% relative uncertainty can still be useful for its order of magnitude), they are not very helpful for exploring or quantifying spectro-temporal relationships between burst properties and can be a hindrance. A subsequent filter removes measurements obtained with only 2 or fewer acceptable arrival times. A more conservative threshold, such as 10 or even 20 or more acceptable arrival times, could be imposed, but this had little effect on our conclusions other than excluding many measurements.

For the datasets analysed here, 1320 measurements were obtained from the arrival times pipeline, and 433 measurements remain after filtering, i.e., about two-thirds of measurements were excluded. This comes after several rounds of review and efforts to obtain accurate measurements for each pulse in the dataset.

The two-thirds measurement exclusion rate indicates a likely bias in our sample towards higher S/N bursts as well as well-resolved components. The large number of dropped measurements in this sample is primarily due to a significant number of very low S/N bursts that subsequently fail the spectral filter applied when determining valid arrival times. Spectro-temporal analysis generally requires a sufficiently high S/N. There is also likely a bias excluding some blended sub-burst components, since these arrival times can be overwhelmed by signal from surrounding components.

\section{Observations and Data}\label{sec:data}

The data used for this study was obtained from previously published observational studies. The majority of the bursts analysed here have not had measurements of their sub-burst slopes performed before, and we also reanalyse the spectro-temporal properties of many bursts from earlier studies. To the best of our knowledge, the ultra-FRBs analysed here represent the first sub-burst slope measurements for these extremely short duration bursts. This section provides a brief summary of the repeating FRBs included in this study, with a particular focus on those exhibiting ultra-FRBs or bursts with microsecond-long durations (specifically, FRB 20121102A, FRB 20220912A, and FRB 20200120E). Additionally, we outline the properties of the bursts analysed and describe the DM strategy employed during the measurements. In general, we try to use the DM of the shortest duration burst in a cohort and apply it to all the bursts when performing measurements. Later, we recalculate the DM of each burst individually and repeat our measurements to assess any major differences. As in earlier studies \citep{Chamma2023,Brown2024}, we process and convert all data from various formats (e.g., \texttt{PSRCHIVE}, \texttt{PSRFITS}, filterbank, etc.) into Python \texttt{numpy} (.npz) format for analysis. The details of this conversion are described here, with reference to previous works where appropriate.

Each burst analysed may have unique inputs to the arrival times pipeline, including the times of sub-bursts, cut placements, mask locations, downsampling factors, and other user-defined options. A measurement script is provided, which lists the options used for each analysed burst and performs the measurements via the arrival times pipeline.

Table \ref{tab:sources} lists the 12 repeating FRB sources analysed in this study. For each source, the table provides the corresponding journal reference for the observational study, the telescope facility used, the observational band covered, and the number of sub-bursts obtained. Additionally, the burst duration range covered by the entire cohort of bursts from a source is included.

\begin{figure*}
    \includegraphics[width=\textwidth]{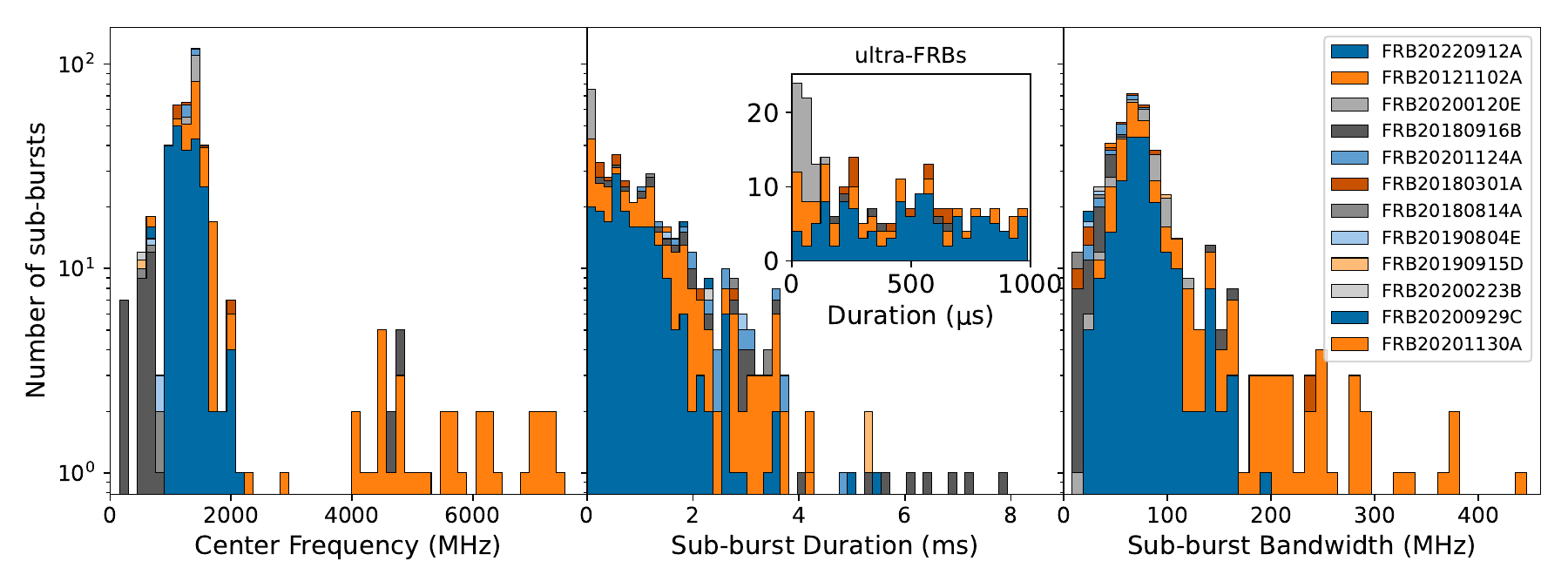}
    \caption{\label{fig:hist}Distribution of burst properties analysed, as measured by the arrival times pipeline. From left to right, the center frequency, sub-burst duration, and sub-burst bandwidth distributions are shown. Bins from each source are stacked. Sources with the most overall bursts (starting with FRB 20220912A) are at the bottom of each bin. Note that the last two colors listed in the legend are repetitions, but these sources only include 4 bursts and their bars are too narrow to see. Inset in the middle panel is an additional plot showing the duration distribution of FRBs shorter than 1 ms with a bin width of 40 $\upmu$s. Sub-bursts here refers to the components of each FRB, each of which have been separated and measured independently.}
\end{figure*}

Figure \ref{fig:hist} displays histograms of the burst properties analysed in this study.

\subsection{FRB Sources}

\begin{table*}
\begin{centering}
\begin{tabular}{lllrrrr}
\toprule 
\textbf{FRB Source} & \textbf{Reference} & \textbf{Telescope} & \textbf{Obs. Band (MHz)} & \textbf{Duration Range} & \textbf{Sub-bursts Measured} & \textbf{$\text{DM}_{\text{applied}}$ (pc/cm$^{3}$)} \tabularnewline
\midrule
\midrule 
FRB 20121102A & \citealt{Hewitt2022} & Arecibo & 1150 -- 1730 & 1.6 $\upmu$s -- 4.2 ms & 21 & 560.5 \tabularnewline
 & \citealt{Li2021} & FAST & 1000 -- 1500 & & 34 & 560.105 \tabularnewline
 & \citealt{Aggarwal2021} & Arecibo & 974 -- 1774 & & 17 & '' \tabularnewline
 & \citealt{Gajjar2018} & GBT & 4000 -- 8000 & & 15 & '' \tabularnewline
 & \citealt{Michilli2018} & Arecibo & 4100 -- 4900 & & 12 & '' \tabularnewline
 & \citealt{Oostrum2020} & WSRT & 1250 -- 1450 & & 12 & '' \tabularnewline
 & \citealt{Scholz2016} & GBT & 1600 -- 2400 & & 4 & '' \tabularnewline
 &  & Arecibo & 1150 -- 1730 & & 1 & '' \tabularnewline
 & \citealt{Snelders2023} & GBT & 3900 -- 9300 & & 7 & '' \tabularnewline
\midrule 
FRB 20220912A  & \citealt{Sheikh2024} & ATA & 900 -- 2334 & 13 $\upmu$s -- 5.5 ms & 135 & 219.356 (see text)\tabularnewline
 & \citealt{Hewitt2023} & NRT & 1230 -- 1742 & & 19 & ''\tabularnewline
 & \citealt{Zhang2023} & FAST & 1000 -- 1500 & & 51 & '' \tabularnewline
\midrule 
FRB 20200120E & \citealt{Nimmo2023} & Effelsberg & 1200 -- 1600 & 13.1 -- 133 $\upmu$s& 28 & 87.7527 \tabularnewline
 & \citealt{Nimmo2022} & Effelsberg & 1200 -- 1600 & & 4 & 87.75 \tabularnewline
\midrule 
FRB 20180916B & \citealt{CHIME2019} & CHIME & 400 -- 800 &  0.18 -- 7.9 ms & 7 & Burst DM \tabularnewline
 & \citealt{Pleunis2021} & LOFAR & 110 -- 188 & & 7 & 348.772 \tabularnewline
 & \citealt{Marthi2020} & uGMRT & 550 -- 750 & & 15 & 348.82 \tabularnewline
 & \citealt{Bethapudi2023} & Effelsberg & 4000 -- 6000 & & 3 & 348.82 \tabularnewline
\midrule 
FRB 20201124A & \citealt{Hilmarsson2021a} & Effelsberg & 1200 -- 1520 & 1 -- 4.9 ms & 16 & 411.60 \tabularnewline
FRB 20180301A & \citealt{Luo2020} & FAST & 1000 -- 1500 & 0.21 -- 2.8 ms & 14 & Burst DM (\textasciitilde 516) \tabularnewline
FRB 20180814A & \citealt{CHIMEFRB2021} & CHIME & 400 -- 800 & 1.25 -- 3.4 ms & 3 & Burst DM (\textasciitilde 189) \tabularnewline
FRB 20200929C & \citealt{CHIME2023} & CHIME & '' & 1.75 -- 2.3 ms & 2 & Burst DM (\textasciitilde 413) \tabularnewline
FRB 20190804E & '' & CHIME & '' & 1.5 -- 2.9 ms & 2 & Burst DM (\textasciitilde 363) \tabularnewline
FRB 20190915D & '' & CHIME & '' & 5.3 ms & 1 & Burst DM (\textasciitilde 489) \tabularnewline
FRB 20200223B & '' & CHIME & '' & 2.3 ms & 1 & Burst DM (\textasciitilde 202) \tabularnewline
FRB 20201130A & '' & CHIME & '' & 1.58 -- 4.3 ms& 2 & Burst DM (\textasciitilde 288) \tabularnewline
\midrule 
 &  &  &  & & \textbf{Total: 433} &  \tabularnewline
\bottomrule
\end{tabular}
\par\end{centering}
\caption{\label{tab:sources}Summary of observations used, including the source, telescope, observational band, duration range, the number of of sub-bursts measured, and the applied DM. As discussed in the text, the DM of a microburst from its respective source is applied to the data for FRB 20121102A (except for the data from \citealt{Hewitt2022}) and FRB 20220912A. In the remaining datasets the DM applied is either the burst DM on a burst-by-burst basis or the DM applied by the authors of the dataset. For datasets that use the burst DM, the approximate DM for that source is listed in parentheses. The duration range is found using measurements from the arrival times pipeline and covers only the bursts with valid measurements from a source.}

\end{table*}

\subsubsection{FRB 20121102A}

FRB 20121102A, the first discovered and one of the most well-observed sources of repeating FRBs, exhibits bursts across a broad range of frequencies, has a candidate 161 day periodicity in its activity cycle, and has been localized to a bright star-forming region on the outskirts of a dwarf galaxy \citep{Spitler2016,Scholz2016,Marcote2017,Bassa2017,Chatterjee2017,Tendulkar2017,Cruces2020}.

\citet{Snelders2023} reported the discovery of 8 bursts from FRB 20121102A with microsecond durations, identified through a reanalysis of 2017 data taken with the Green Bank Telescope (GBT). These bursts are orders of magnitude shorter than other known FRBs. The data and processing scripts were made available via a Zenodo link and were adapted to produce .npz files for the arrival times pipeline. 

In \citet{Chamma2023}, we analysed a subset of the bursts observed by \citet{Michilli2018,Gajjar2018,Oostrum2020,Aggarwal2021} and \citet{Li2021}. Waterfalls from these sources were obtained in ASCII, \texttt{PSRFITS}, or \texttt{filterbank} format, and underwent downsampling, dedispersion, and other processing steps before spectro-temporal analysis. A detailed description of how the waterfall data were prepared and converted to Python .npz files can be found in Section 3 of \citet{Chamma2023}. These same files were used in re-analysing the data with the arrival times pipeline for this study. We note that burst 11M from \citet{Gajjar2018} potentially consists of two microshots (noticeable at a DM of 560.105 pc/cm$^3$) that were previously missed. However, these microshots had too low of a S/N to be reliably measured. 

Data from \citet{Hewitt2022} included 478 bursts observed during a burst storm in 2016. In \citet{Brown2024}, 24 of these bursts were analysed using \textsc{Frbgui} \citep{Chamma2023}. We re-analyzed the entire dataset and obtained measurements for 46 sub-bursts, representing a small portion of the total number of bursts (as listed in Table \ref{tab:sources}). There are several reasons why measurements could not be obtained from the majority of the bursts. These are typically due to very low S/N, high uncertainties arising from the quality of the waterfall data and measurements, as well as radio frequency interference (RFI) in the waterfall.

The final dataset for this source included the six bursts observed by \citet{Scholz2016}, which were already provided in .npz format. These are the only bursts in the 2 GHz range for FRB 20121102A that the authors are aware of.

For this dataset we investigated applying the DM of one of the shortest bursts from the source (in this case 560.105 pc/cm$^3$ from the ultra-FRB B30 of \citealt{Snelders2023}) to the majority of the data. This approach was chosen because the DM of the shortest duration burst is likely to have high precision due to its sharp profile and high frequency. \citet{Jahns2023} also recommend this step when analysing bursts emitted within weeks of each other, as their results suggest that short-term DM variations are caused by intra-burst drift. The data from \citet{Hewitt2022} were not included in this step, and these data were analysed with their applied DM of 560.5 pc/cm$^3$. We hypothesized that the DM of the shortest duration burst could be a highly precise measurement of a source's DM, which would be indicated by an improved agreement with fits to the sub-burst slope law. We did indeed observe slight improvements in the goodness of fit compared to using each burst's individual DM. In Sections \ref{subsec:burstdm} and \ref{subsec:indivdms}, we detail the process of applying each burst's individual DM including, recalculating DMs when necessary. The results of these DM investigations are discussed further in Appendix \ref{subsec:dmopt}.

\subsubsection{FRB 20220912A}

FRB 20220912A is a recently discovered and highly active repeating source, capable of emitting hundreds of bursts per hour during periods of activity \citep[][peak event rate of 390 hr$^{-1}$ with a 90\% fluence threshold of $\sim$0.014 Jy$\cdot$ms]{Zhang2023}. Some of these bursts are exceptionally bright and display dense microshot structures, as observed by \citet{Hewitt2023}. The source has been localized to a galaxy with a low host contribution to the DM along the line of sight \citep{Ravi2023}.

The data analysed from \citet{Hewitt2023} were prepared using the scripts provided in that paper's reproduction package\footnote{\url{https://zenodo.org/records/10552561}}. Since some microshots in this dataset contained over 40 components with varying degrees of blending, we initially used CHIME's \texttt{fitburst} package \citep{Fonseca2023} to construct a spectral model of the burst with the appropriate number of components, and then extracted the model components for measurement using the arrival times pipeline. Although we were eventually able to produce a reasonably accurate model with the correct number of components, the extracted components appeared too synthetic, and the drifting information within sub-bursts was lost, as the model components were perfectly vertical. Consequently, we reverted to using the feature within the arrival times pipeline that allows for manual selection of each component's location, along with manually placing cuts between components to mitigate the effects of blending (see Appendix \ref{app:microshot} for more details).

We obtained 35 burst waterfalls from \citet{Sheikh2024} as \texttt{numpy} npz files and PSRFITS archives, however, the npz files were sufficient and we used those directly. These bursts were observed by the Allen Telescope Array (ATA), which covered a frequency band of 900 – 2334 MHz across multiple tuning configurations. Our analysis also includes over 60 additional FRBs, many with complicated series of components, that were discovered after publication \citep{Sheikh2024a}. Almost all these additional FRBs exhibited multiple components and ended up forming the majority of measurements obtained from this dataset.

Burst waterfalls from figures 5 and 9 of \citet{Zhang2023} were obtained as raw Python arrays at a DM of 220 pc/cm$^3$, and we added metadata such as the frequency axis and DM for use in analysis. While that study detected 1076 bursts, whose corresponding data is publicly available, only those published in the figures were available at the time of analysis and we leave the rest of the dataset for future work. Additionally, we optionally centered and normalized each channel in each waterfall by the off-pulse mean and standard deviation, as this improved the visual clarity of the burst.

We applied the DM of the microshot forest B1, as determined by \citet{Hewitt2023} (219.356 pc/cm$^3$), to most bursts from this source. For the two additional, less complex microshot forests (B2 and B3) from \citet{Hewitt2023}, we chose to retain their original DMs of 219.375 and 219.8 pc/cm$^3$, respectively, due to the effort involved in determining those DMs.

\subsubsection{FRB 20200120E}

FRB 20200120E is a repeating FRB source, currently the closest discovered to our galaxy. Unlike other sources typically found in younger star-forming regions, it is localised to a globular cluster \citep[e.g.][]{Bhardwaj2021,Kirsten2022,Nimmo2023}. Its burst morphology is also exceptional, exhibiting structure down to the nano-second scale \citep{Nimmo2022}.

We accessed the five bursts reported in \citet{Nimmo2022} via the Zenodo link provided, which included archive files of the bursts generated from filterbank data. Using \texttt{pypulse}, we loaded the archives, dedispersed them to the reported DM of 87.75 pc/cm$^3$ and downsampled the frequency channels by a factor of 8. 

For the approximately 60 bursts reported in \citet{Nimmo2023}, the data were also available in filterbank format again through a Zenodo link. We used the \texttt{your} package to load the data and incoherently dedispersed all bursts to the reported DM of 87.7527 pc/cm$^3$. The data were cropped and normalized by channel to improve burst visibility. Also to boost visibility and S/N, we downsampled the bursts in frequency by a factor of 8 and in time by a factor of 2, except for three bursts detected with a pulsar backend, which had a longer time resolution and were only downsampled in frequency. Additionally, masking was applied to channels affected by RFI across all waterfalls.

\subsubsection{Remaining sources}

The remaining data were collected from repeating FRB sources as part of the analysis conducted by \citet{Brown2024}. We summarize relevant information regarding these sources here and refer the reader to Section 2.2 of \citet{Brown2024} and Table \ref{tab:sources} for more details. 

FRB 20180916B is a repeating source with a periodic activity cycle of approximately 16 days, exhibiting chromatic activity over the course of its cycle. It is located in a spiral galaxy 149 Mpc away \citep[e.g.,][]{CHIME2020b,Tendulkar2021,PastorMarazuela2021}. Data for this source extend to the lowest frequencies observed from FRBs, including the 110 MHz detections reported by \citet{Pleunis2021}.

Repeaters FRB 20180301A and FRB 20201124A are well-observed sources, with studies frequently focusing on the polarization properties of their emission. Here, we re-analyse bursts from these sources as presented in \citet{Luo2020} and \citet{Hilmarsson2021a}, respectively.

Data for sources FRB 20180814A, FRB 20190804E, FRB 20190915D, FRB 20200223B, FRB 20200929C, and FRB 20201130A are accessed from the CHIME/FRB burst catalogs \citep{CHIME2019,CHIMEFRB2021,CHIME2023}.

\subsection{Burst DM Measurements}\label{subsec:burstdm}

As mentioned, to assess any major differences in our results due to the choice of DM, we repeat our measurements at the individually determined DM of each burst. We repeated all spectro-temporal measurements at the burst DM reported by the original authors, if there was one. For bursts without an individually calculated DM, we used \texttt{DM\_phase}\footnote{\url{https://github.com/danielemichilli/DM_phase}} \citep{CHIME2019a,Seymour2019} to recalculate the DM using an algorithm that preserves structure by maximizing coherent power instead of S/N, and repeated the measurement at that DM. The datasets whose burst DMs we recomputed were those from \citet{Gajjar2018,Michilli2018,Oostrum2020,Scholz2016,Snelders2023,Hewitt2022,Zhang2023,Nimmo2022,Nimmo2023,Pleunis2021a,Marthi2020,Bethapudi2023}. For most of the FRBs we recomputed DMs for, we searched for optimal DMs in a range of $\pm$ 40 pc/cm$^3$ from the applied DM in steps of 0.1 pc/cm$^3$. For shorter duration FRBs (namely those from \citealt{Snelders2023} and \citealt{Nimmo2022,Nimmo2023}, as well as bursts from FRB 20180916B), a finer grid was needed over a smaller range, so we searched DMs $\pm$ 10 pc/cm$^3$ from the applied DM in steps of 0.01 pc/cm$^3$. For the remaining datasets that reported individual burst DMs, we repeated our measurements at those DMs.

% \subsubsection{FRB 20180916B}
% \subsubsection{FRB 20201124A}
% \subsubsection{FRB 20180301A}
% \subsubsection{FRB 20180814A}
% \subsubsection{FRB 20200929C}
% \subsubsection{FRB 20190804E}
% \subsubsection{FRB 20190915D}
% \subsubsection{FRB 20200223B}
% \subsubsection{FRB 20201130A}

%%%%%%%%%%%%%%%
% * Describe arrival times method and capabilities
% * Describe data and bursts measured
    % * DM distribution with time
% * Describe DM selection/bootstrapping process

\section{Results}\label{sec:results}

Our analysis provides spectro-temporal measurements for each burst examined, including the center frequency ($\nu_0$), duration ($\sigma_t$), bandwidth ($\sigma_\nu$), and (inverse) sub-burst slope/intra-burst drift ($\text{d}t/\text{d}\nu$), along with their uncertainties. 

\begin{figure*}
    \includegraphics[width=\textwidth]{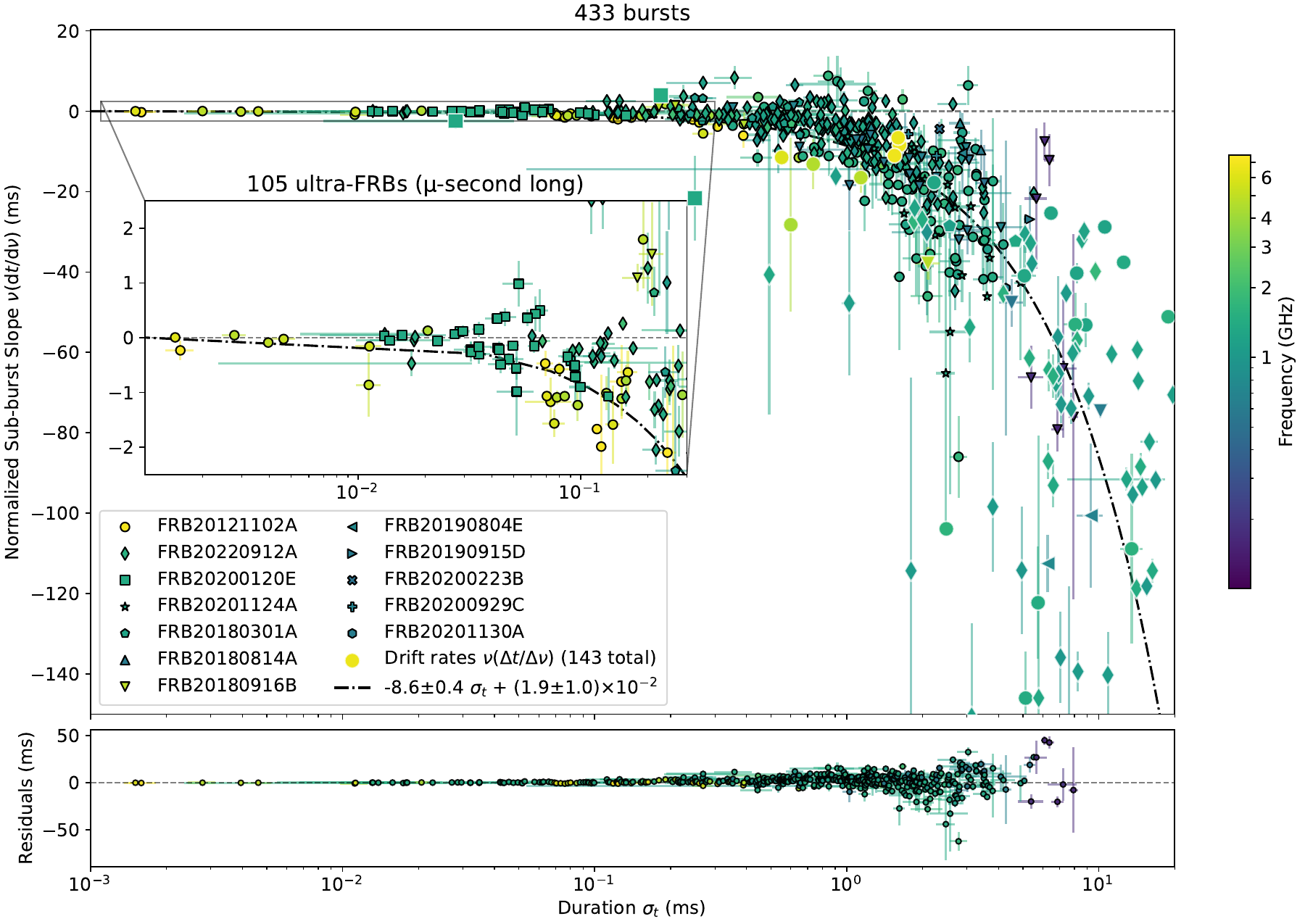}
    \caption{\label{fig:nudtdnu}Plot of the inverse normalized sub-burst slope, $\nu (\text{d}t/\text{d}\nu)$, versus the sub-burst duration, $\sigma_t$, for a cohort of FRBs from multiple repeating sources. All measurements were obtained using the arrival times pipeline. The duration axis is displayed on a logarithmic scale. The inset panel provides a zoomed view of ultra-FRBs with durations below our selected upper limit of 300 $\upmu$s. The marker shape denotes the source of the burst, while bursts are colored by frequency, ranging from around 100 MHz to nearly 7.5 GHz. The dashed-dotted line represents a general linear fit to the data from FRB 20121102A, the source with the most extensive set of observations. Bursts shown here are generally measured at the DM of the shortest duration burst in their cohort; see the text and Table \ref{tab:sources} for additional details on the DMs used for measurements. The residuals between all bursts and the plotted fit are shown in the bottom panel. Overall, we observe good agreement with the linear fit across all sources within the uncertainties, consistent with previous analyses. Drift rate measurements obtained from waterfalls with multiple resolved components are overlaid with larger markers and a white border. These measurements generally align with the fit for sub-burst slopes, but they exhibit more outliers. Some drift rates fall outside the chosen axis limits. See Figure \ref{fig:extradata} for a version of this plot with measurements from other studies.}
\end{figure*}

In this section, we explore the correlations between each of the parameters measured in order to verify those observed in \citet{Chamma2023} and \citet{Brown2024} (among others) and to leverage the expanded parameter space afforded by our measurements from ultra-FRBs. We focus on the relation between the (inverse) sub-burst-slope and duration in Figure \ref{fig:nudtdnu} and present fits to this relationship for each repeating FRB source in Table \ref{tab:fits}. Additional correlations are shown in Figure \ref{fig:corner}, with points colored by duration to highlight the parameter space occupied by ultra-FRBs. We will also contextualize the observed spectro-temporal correlations in light of previous findings. 

Section \ref{sec:drifts} summarizes the drift rate measurements obtained and their relationship with duration. Section \ref{sec:methodcmpr} presents the results of comparison tests between the arrival times and Gaussian methods, highlighting areas of agreement, and conditions under which inconsistencies may arise.

%Finally, Appendix \ref{app:microshot} demonstrates the capabilities of the arrival times pipeline, detailing the measurement process and displaying the measurements obtained from a microshot forest with over 40 components.

% Drift rates?

% * Show some examples of measurements

% * Show ultra-FRBs analysed (frb20121102a, frb20220912a, frb20200120e)

% * Show microshot forests from hewitt, sheikh, and zhang to demonstrate capability of arrival times pipeline in analysing these bursts

% * Table of fit results (both slope and intercept) broken down by source

% * Briefly describe results of DM selection/bootstrapping process

\begin{table*}
\begin{centering}
\begin{tabular}{lcccr}
\toprule 
\textbf{FRB Source} & \multicolumn{2}{c}{\textbf{Fit} $\nu\cdot\text{d}t/\text{d\ensuremath{\nu}}=a\sigma_{t}+b$} & $\chi^{2}_{\text{red}}$ & \textbf{\# Bursts}\tabularnewline
\midrule 
 & $a$ (unitless) & $b$ (ms) & \tabularnewline
\midrule
% \midrule 
FRB20121102A & -8.6$\pm$0.4 & (1.9$\pm$1.0)$\times 10^{-2}$	& 18.40 & 122 \tabularnewline
FRB20220912A & -7.4$\pm$0.3 & (3.1$\pm$0.6)$\times 10^{-1}$	& 68.75 & 205 \tabularnewline
FRB20200120E & -5.4$\pm$1.6 & (1.1$\pm$0.5)$\times 10^{-1}$	& 4.90 & 32 \tabularnewline
FRB20180916B & -8.9$\pm$1.3 & 4.0$\pm$0.6 & 13.27 & 32 \tabularnewline
FRB20201124A & (-1.7$\pm$0.2)$\times 10^{1}$ & (1.7$\pm$0.4)$\times 10^{1}$	& 4.38 & 16 \tabularnewline
FRB20180301A & (-1.2$\pm$0.4)$\times 10^{1}$ & 4.4$\pm$1.7 & 11.40 & 14 \tabularnewline
% sources with 3 bursts or less are not included
% \midrule 
 % &  &  &  & \textbf{Total: 384}\tabularnewline
\bottomrule
\end{tabular}
\par\end{centering}
\caption{\label{tab:fits}Results of linear fits to sub-burst slope law for each source with more than a few measured bursts. The fit follows the form $\nu(\text{d}t/\text{d}\nu) = a\sigma_t + b$, where $\text{d}t/\text{d}\nu$ is the inverse sub-burst slope and $\sigma_t$ is the sub-burst duration. Uncertainties are at the 1$\sigma$ level.}
\end{table*}

\subsection{Spectro-temporal relationships}\label{sec:relationships}

\subsubsection{Sub-burst slope vs. duration relation}
The sub-burst slope relationship for the measured FRBs is shown in Figure \ref{fig:nudtdnu}, which displays the inverse normalized sub-burst slope $\nu (\text{d}t/\text{d}\nu)$ versus sub-burst duration. The inset panel highlights measurements from ultra-FRBs, defined here as bursts with durations below an arbitrarily chosen cutoff of 300 $\upmu$s. Data points are colored by frequency. In this plot, bursts with a near vertical orientation in their waterfall have a value of $\text{d}t/\text{d}\nu$ approaching zero. The ultra-FRBs exhibit a strong clustering around zero, with fluctuations diminishing as the duration decreases. 

The predicted relationship between sub-burst slope and duration for FRBs is provided in Equation 7 of \citet{Rajabi2020}. In the formalism used here, this relationship is inverted, taking the following linear form 
\begin{equation}\label{eq:slopelaw}
    \nu\frac{\text{d}t}{\text{d}\nu}=-A\sigma_{t},
\end{equation}
where $A$ is a constant. To evaluate the agreement with this predicted relationship, we fit the data using the general linear form $\nu(\text{d}t/\text{d}\nu)=a\sigma_{t} + b$, applying orthogonal distance regression \citep{Boggs1990} via the \texttt{scipy.odr} package. This method incorporates uncertainties in both the dependent and independent variables when determining the best fit. 

Table \ref{tab:fits} presents the results of these fits for each FRB source analysed. The fit for FRB 20121102A, which includes 149 bursts measured and covers the widest range of frequencies and durations, yields $a=-8.6(4)$ and $b=0.019(10)$ ms. The uncertainties here and throughout (including figures) are quoted at the 1$\sigma$ level. This fit is shown as the black dash-dot line in Figure \ref{fig:nudtdnu}. Note that the curved appearance of the fit line results from the logarithmic scale used on the duration axis. We check the difference in fit when using the average channel duration instead of the integrated time-series duration. With that duration, the fit changes to $a=-14(1)$ and $b=0.020(14)$ ms. We note however that the residuals are larger in this case and tend to larger positive values for longer duration bursts. Due to the general form of the fit, agreement with Equation \ref{eq:slopelaw} is qualitatively captured by the magnitude of $b$ (as well as $\chi^2_\text{red}$), where a value near zero indicates good agreement and larger values suggest otherwise. Assuming that $b$ must be zero (i.e., that the sub-burst relation holds true) allows $b$ to serve as an indicator of data quality. A large $b$ may result from a limited variety or incomplete sample of bursts in terms of frequency and/or duration. This trend is evident in Table \ref{tab:fits} for the remaining sources, where $a$ fluctuates around the value found for FRB 20121102A, while $b$ generally increases significantly as the number of available bursts decreases. FRB 20220912A challenges this trend since $b$ is large despite the large number of bursts, however this may be due to an actual deviation from the sub-burst slope law for this source.

We can compare our result of $a = -8.6(4)$ for FRB 20121102A with previous analyses. In \citet{Brown2024}, using measurements from 65 sub-bursts, a value of $A = 0.076(4)$ was obtained for FRB 20121102A using the inverse form of Equation \ref{eq:slopelaw}. Converting to our formalism gives $a_1 = 1/A = 13.2(7)$. Similarly, \citet{Chamma2023} found $A = -0.113(3)$ for FRB 20121102A using 167 bursts\footnote{The lower number of bursts used here is partly due to what constitutes a usable measurement when using arrival times versus when using the ACF/Gaussian methods. For example, requiring the S/N $>3\sigma$ in each channel affects which waterfalls end up being measurable.}. Converting that value gives $a_2 = -8.9(2)$, equal to the result found here within uncertainties. In \citet{Jahns2023}, approximately half of the 849 bursts observed from FRB 20121102A in the 1150 -- 1730 MHz band were analysed for intra-burst drift using a 2D Gaussian fit that directly included the drift as a parameter $d_t$. This measurement of $d_t$ has the same dimensions as $\text{d}t/\text{d}\nu$ used here, describing the same property of a burst, and can therefore be directly compared \citep[App. A of][]{Jahns2023}. When performing a linear fit of the form $d_t = b\sigma_t + c$, \citet{Jahns2023} found $b = -0.00862$ MHz$^{-1}$ and $c = 0.00171$ ms/MHz. Converting to the present formalism requires multiplying by their mean burst frequency of $\sim$1450 MHz, yielding $a_3 = -12.5$. The three fit values for the sub-burst slope relation of FRB 20121012A from earlier studies--$a_1, a_2, a_3$--are therefore quite close, with $a_2$ matching the value of $a$ obtained here, despite the significant methodological differences. For example, \citet{Chamma2023} and \citet{Brown2024} fit a 2D Gaussian to the ACF of the burst waterfall, while \citet{Jahns2023} used a differently parametrized 2D Gaussian fit directly to the burst waterfall. Slight differences in the resulting fit values can be due to the smaller number of bursts used, as in the cases of $a_1$ and $a_2$, or from the limited frequency range in the case of $a_3$, as well as the absence of measurements from ultra-FRBs until now. Other differences may stem from variations in the definition of sub-burst duration. Despite these factors, we observe close agreement between the earlier fit values of the sub-burst slope relation for FRB 20121102A and the value obtained here.

\subsubsection{Relations between $\nu_0$, $\sigma_t$, $\sigma_\nu$ and $\text{d}t/\text{d}\nu$}\label{subsubsec:relations}

\begin{figure*}
    \includegraphics[width=\textwidth]{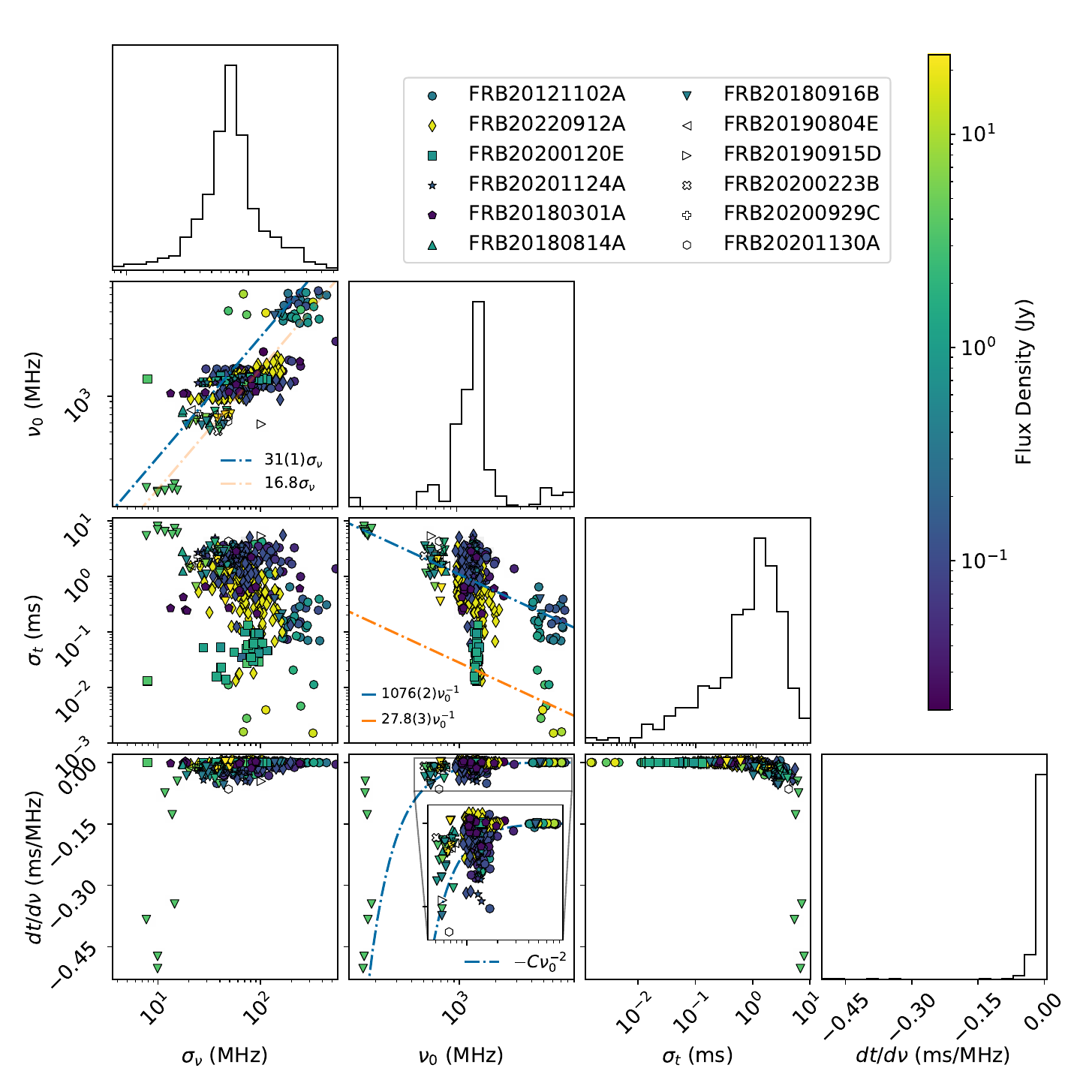}
    \caption{\label{fig:corner}Corner plot of burst properties measured from 12 repeating FRB sources. Marker shape denotes the FRB source. Points are colored by their reported flux density, as indicated by the color bar. Note that some points are colored by the average flux density, as described in the text. Uncolored points have missing flux densities. The strongest correlations are seen in the three $\text{d}t/\text{d}\nu$ plots along the bottom row. The blue fit line shown in the $\text{d}t/\text{d}\nu$--$\nu_0$ plot is the fit found from FRB 20121102A data reported in \citet{Chamma2023}. The blue and orange fit lines in the $\sigma_t$--$\nu_0$ plot are fit to bursts with duration greater and less than 300 $\upmu$s. See the text for details.}
\end{figure*}

Additional correlations between the spectro-temporal properties of the analysed FRBs are presented in the corner plot in Figure \ref{fig:corner}. The burst parameters examined are $\nu_0$, $\sigma_t$, $\sigma_\nu$, and $\text{d}t/\text{d}\nu$. In that figure, marker shape indicates the source, while point color represents the flux density. For the datasets from \citet{Li2021}, \citet{Zhang2023} and \citet{Sheikh2024}, we could not fully confirm the mapping of burst properties to burst filenames to accurately assign each flux density, and therefore used the average flux density to assign the color. For uncolored points, we did not find a reported flux density.

The strongest correlations are those observed with $\text{d}t/\text{d}\nu$, as shown in the bottom row. The $\sigma_t$--$\sigma_\nu$ plot shows little to no correlation, while the $\sigma_t$--$\nu_0$ plot shows some evidence of correlation, though with significant statistical scatter in the measurements.

The $\nu_0$--$\sigma_\nu$ plot appears to exhibit a linear correlation, as was found with previous results obtained over broad frequency ranges but with fewer bursts \citep{Houde2019,Chamma2023,Brown2024}; however, there again remains significant statistical scatter. Additionally, the higher frequency bursts shown are observed with a larger observing bandwidth. The observing bandwidth may therefore be biasing this result. We fit a line to the complete set of measurements and obtain $\nu_0 = 31(1) \sigma_\nu$. Previous values for this correlation are around $\sqrt{8\ln{2}}/0.14\simeq16.8$, after inverting and scaling from the FWHM for comparison here. Despite the present value being nearly double, we note that, because of the spread in measurements, either fit could be visually attributed to the data.

Two fits are overlaid on the $\sigma_t$--$\nu_0$ plot of the form $\sigma_t \propto \nu_0^{-1}$. The blue and orange lines are fit to all bursts with durations longer and shorter than 300 $\upmu$s, respectively. We find $\sigma_t \propto1076(2)$ [ms$\cdot$MHz] $\nu_0^{-1}$ for the longer FRBs, and $\sigma_t \propto 27.8(3)$ [ms$\cdot$MHz] $\nu_0^{-1}$ for the ultra-FRBs. Despite the statistical scatter in measurements, the blue line reflects well the trend of measurements, especially the very low frequency bursts from FRB 20180916B. The fit to the ultra-FRBs (orange) also describes the shortest duration bursts well. Bursts at around 1.4 GHz, especially from the sources FRB 20200120E and FRB 20220912A, connect the duration gap between the two fits, suggesting a continuum may exist. In that case, the blue and orange fits may represent opposite ends of an envelope. The functional dependence is the prediction of the TRDM and consistent with results reported in \citet{Chamma2023}. There is room for other interpretations and functional forms that may describe this data (such as $\sigma_t \propto \nu_0^{-2}$). However, we focus here on the context that the TRDM provides and follow predictions made by \citet{Kumar2024}, which investigates modifications to the TRDM at short durations and under various propagation effects. In \citet{Kumar2024}, bursts that obey the sub-burst slope law and vary in duration by several orders of magnitude can show the sort of separation reflected in the two fits shown.

% Two fits are overlaid on the $\sigma_t$--$\nu_0$ plot. The blue fit corresponds to the correlation found in \citet{Chamma2023}, which follows the form $C\nu_0^{-1}$, where $C=1474$ (ms$\cdot$MHz). This fit no longer accurately describes the data, primarily due to the addition of the ultra-FRBs (yellowish points). While the relation $\sigma_t \propto \nu_0^{-1}$ may still be applicable with a different scaling, there is significant statistical scatter in the burst measurements at each frequency, and it appears equally plausible that $\sigma_t \propto \nu_0^{-2}$ (shown in orange) could be an appropriate model, or that these two parameters are uncorrelated. The inclusion of ultra-FRBs from each of the three ultra-FRB sources (FRB 20121102A, FRB 20220912A, and FRB 20200120E) has weakened the connection between $\sigma_t$ and $\nu_0$. However, an upper envelope may still exist, given the absence of longer-duration bursts observed at higher frequencies (approximately 4–8 GHz).

The blue fit in the $\text{d}t/\text{d}\nu$--$\nu_0$ plot, which follows the form $C\nu_0^{-2}$, appears to support the relationship previously reported in \citet{Chamma2023} between these two parameters. Using that earlier fit result, we take $ C = -1/(6.1\times10^{-5}) \simeq 16340$ ms$\cdot$MHz, inverted for our results here. While caution should be exercised in applying this fit to data from multiple sources, it is evident that the fit qualitatively represents the data reasonably well, including the very low-frequency data from FRB 20180916B (though 3 of the 7 LOFAR points lie in the top left of the plot and do appear to be outliers). In the inset, the fit passes neatly through the points representing ultra-FRBs, which were not available when $C$ was found. Additionally, we observe significant statistical scatter around 1500 MHz, a characteristic common to data from all the repeating sources included. 

Another notable correlation, specifically between $\text{d}t/\text{d}\nu$ and $\sigma_\nu$, was also observed in \citet{Chamma2023}. We detect this correlation again here and note that, due to measurement noise, the data could potentially be described by relationships of several forms (e.g. $\text{d}t/\text{d}\nu \propto \sigma_\nu^{-1}$ to $\sigma_\nu^{-3}$). A conclusive form cannot be inferred from the data. The relationship between $\text{d}t/\text{d}\nu$ and $\sigma_t$ is addressed in the previous section, where $\text{d}t/\text{d}\nu$ is multiplied by frequency to normalize it across datasets.
% Because of differences in the definition for bandwidth used, the fit value in \citet{Chamma2023} will not map straightforwardly the measurements obtained here.

The interpretation of these relationships within the context of the TRDM, as well as the effect of dispersion and scattering on these relationships, is discussed in Section \ref{sec:discussion}. 

\subsubsection{Changes when measuring at individual burst DMs}\label{subsec:indivdms}

We summarize here the differences on our results when the analysis is repeated at each burst's individually calculated DM (Section \ref{subsec:burstdm}). While major changes can be seen to the values of d$t$/d$\nu$ measured burst-to-burst, the general relationships and trends between properties are the same, though with the caveat that the fit parameters can change significantly. 

Generally, we observe an expected shift of slope measurements upwards towards positive slopes. After applying the same measurement filters applied to the earlier measurements (Section \ref{subsec:filters}), 325 sub-burst measurements remain (shown in Figures \ref{fig:nudtdnu_burstdm} and \ref{fig:corner_burstdm}). The reduction in burst count is primarily due to the increase in uncertainty for values of d$t$/d$\nu$ that are close to or overlap with 0 ms/MHz. Since the burst DM is found using an algorithm that generally aligns bursts vertically, this situation is more common and we observe a corresponding decrease in usable measurements. Nonetheless, we still observe a downwards trend with increasing duration consistent with the sub-burst slope relation. The fit parameters obtained for the sub-burst slope relation for bursts from FRB 20121102A was $a=-3.5(5)$ and $b=(8.7 \pm 14.2) \times 10^{-3}$ ms, as well as a larger reduced-$\chi^2 \simeq 41$. For the remaining sources, the fit parameters found showed similar differences, with the reduced-$\chi^2$ either similar or much larger compared to those tabulated in Table \ref{tab:fits}.

For the other spectro-temporal relationships examined (namely those in Figure \ref{fig:corner}), the $\nu_0-\sigma_\nu$ fit still shows a linear trend but changed from $\nu_0=31(1)\sigma_\nu$ to $\nu_0=20(0)\sigma_\nu$. The $\sigma_t-\nu_0$ trend is visually the same and the fit changed from $\sigma_t=1076(2)\nu_0^{-1}$ to $\sigma_t=970(2)\nu_0^{-1}$. The remaining trends look visually similar with the main difference being three low frequency points from LOFAR (FRB 20180916B) now having positive d$t$/d$\nu$ values, and thus appearing more significant outliers from the remaining measurements. 

\subsection{Drift rates}\label{sec:drifts}

As explained in Section \ref{sec:methods}, FRBs with multiple components have their drift rates measured by fitting a line through the central frequency and time of each component (see the bottom panel of Figure \ref{fig:arrmethod} for an example). 

After reviewing the measurements and excluding those with relative uncertainties greater than 100\% (since even just having the order of magnitude can be informative), a total of 143 drift rates were measured across our burst sample. The magnitudes of the drift rates, $|\Delta t/\Delta \nu|$, span a range of approximately 0.001 -- 7.5 ms/MHz, covering about four orders of magnitude. Thirteen of these appear to be plausible positive `happy trombones', which could be due to a physical process or simply a coincidence. In the TRDM, the drift rate sign is determined by the rate of change of the center frequency with the delay time \citep{Rajabi2020}.

To validate our measurements against the ACF/Gaussian methods, we checked that our drift rates align with previous measurements for bursts that have been studied before. For example, for burst B1 from \citet{Hewitt2023}, which contains approximately 40 components with widely varying durations, we measure a drift rate of $-0.134\pm 0.006$ ms/MHz, or $-7.5 \pm 0.3$ MHz/ms. The drift rate reported for this burst using a 2D Gaussian fit to the ACF was $-8 \pm 3$ MHz/ms, which is consistent within the uncertainties. Additionally, \citet{Konijn2024} analysed a superset of the bursts observed by \citet{Hewitt2023} and found a drift rate of $-7.3 \pm 0.2$ MHz/ms for this burst (B411 in their naming scheme), which agrees even more closely with our measurement.

\begin{figure}
    \includegraphics[width=\columnwidth]{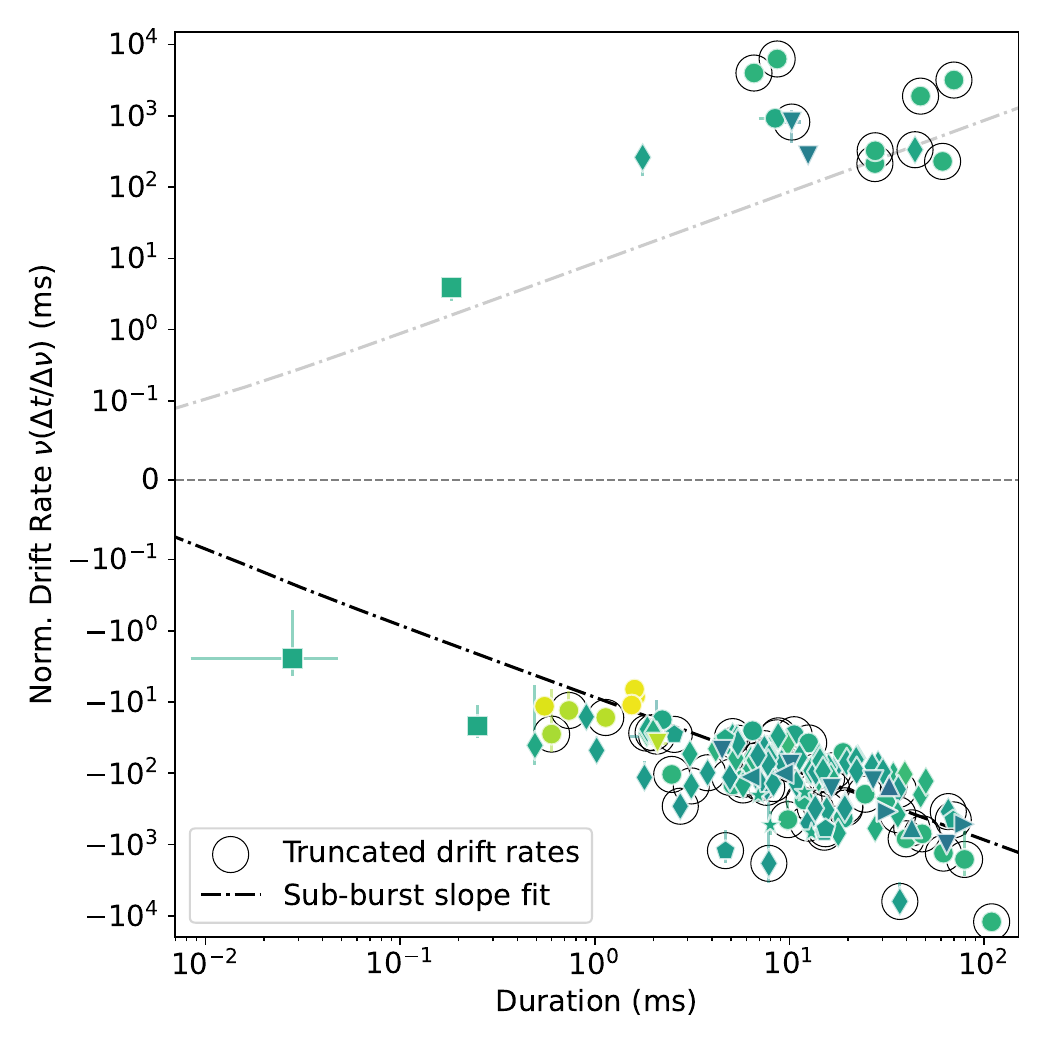}
    \caption{\label{fig:drift}Multi-component drift rate measurements of 143 FRBs, normalized by the center frequency, plotted against duration, are shown analogous to Figure \ref{fig:nudtdnu}, but with a semilog scale on the y-axis. Colors and markers are the same as before. The dashed line represents the corresponding fit found in Figure \ref{fig:nudtdnu} for the sub-burst slope measurements. Its negative is shown with the faded line. The circled drift rates (71 in total) represent drift rates measurements from waterfalls where at least one burst component is truncated by the observing band. There is good agreement between the drift rates and the relationship found for the sub-burst slopes. Despite being truncated, the circled drift rates are not significant outliers from the fully captured drift rates.}
\end{figure}

Figure \ref{fig:drift} shows the drift rate measurements, each multiplied by the mean frequency of their underlying components, plotted against their duration. Overlaid is the fit obtained for the analogous plot of the sub-burst slopes in Figure \ref{fig:nudtdnu}. We observe good agreement between this fit and the drift rate trend, supporting the findings in \citet{Jahns2023} and \citet{Brown2024} that these two measurements follow similar trends. Interestingly, the positive drifts approximately follow the negative of this relationship, shown with the faded line, though the small number of points makes it difficult to conclude for certain. This suggests a connection with the negative drifts and the existence of a sign change \citep{Rajabi2020}, which can explain this aspect of the positive drifts. To double check this result, we repeated all our drift rate measurements using the ACF method. We obtained 132 drift rates from our sample in this way (shown in Figure \ref{fig:driftacf}). Using the same durations, we saw similar trends, despite some differences in measured drifts. With the ACF drift rates, the values move above the line whereas they were below in Figure \ref{fig:drift}. The positive drift rates also exhibit a larger spread in measurements. We note that truncated drift rates are also potentially inaccurate using the ACF method (just as they can be with the arrival times method). Truncated signals can lead to a rotation of the ACF, changing the measured drift rate. Additionally, as the separation between components becomes greater, multiple distinct features can arise in the ACF. This makes the ACF drift rate more challenging and ambiguous to determine as the model may not fit these features well. In this latter case, the arrival times method offers advantages over the ACF method for drift rates.

The agreement between the sub-burst slope relation and drift rate measurements can potentially be explained within the context of the TRDM as resulting from minimal or nonexistent variations in the time between a trigger and FRB emission, as discussed in Section 3.1 of \citet{Chamma2021}. However, as noted by previous authors, this remains a non-trivial and perhaps unexpected result that warrants deeper study and further validation. 

% * run Victor's bursts too?

% * there's a burst that moved in FRB 20121102A after, noticed adding the drift rates stuff. try diffing the spreadsheets. At least two points one from gajjar and one from michilli appear. Review those measurements

% * Update fit, ultra frbs table, and num bursts if needed

\subsection{Method Comparison}\label{sec:methodcmpr}

In order to assess the differences between the arrival times pipeline and Gaussian methods for spectro-temporal measurements, we selected a subset of bursts (partly based on the simplicity of measurement) and repeated their measurements using each method. We found reasonable agreement between measurement values, with similar precisions when using the arrival times pipeline. The ACF method shows the best precision of all the methods, across all duration scales.

\begin{figure}
    \includegraphics[width=\columnwidth]{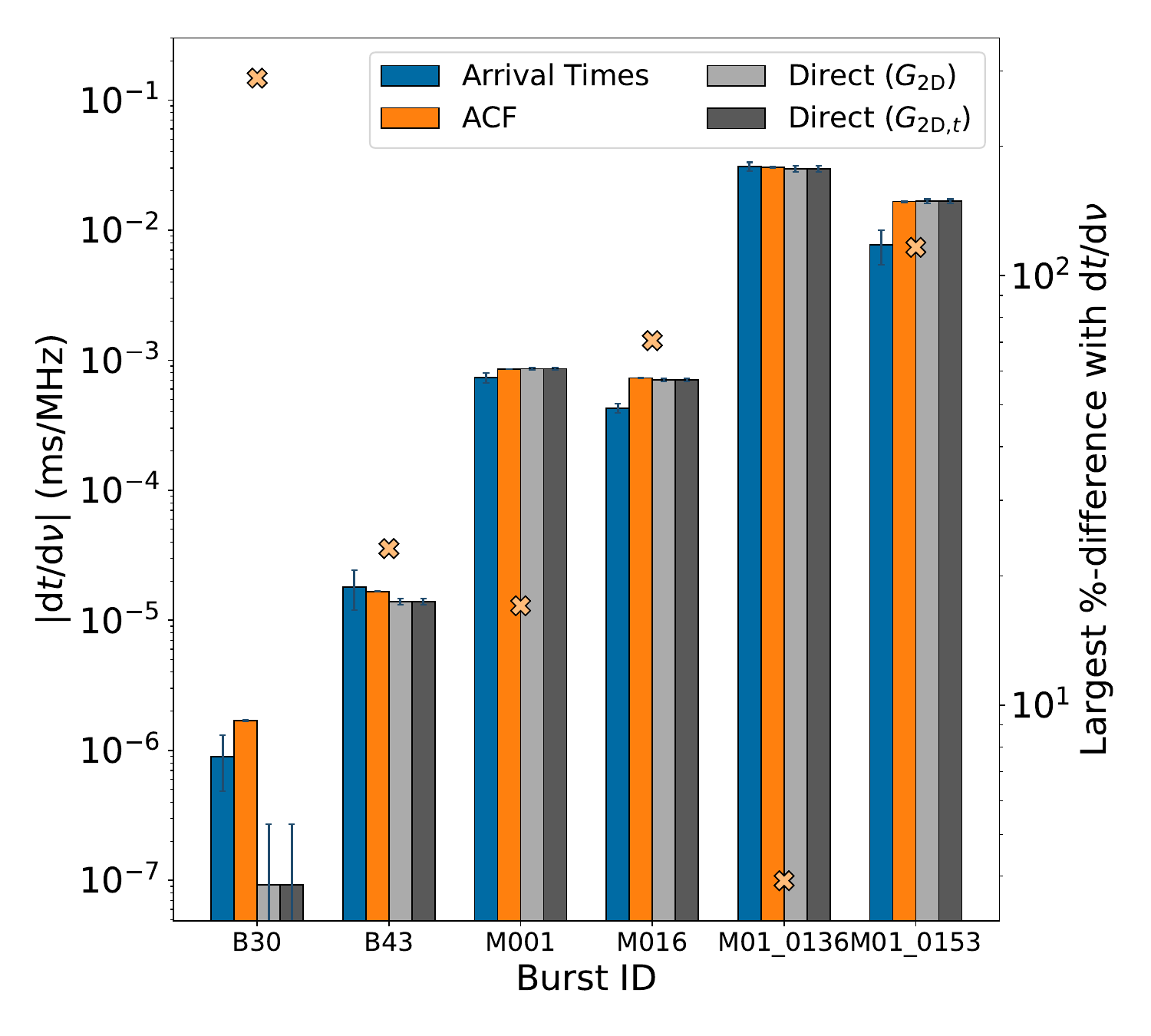}
    \caption{\label{fig:methcompare}Comparison of measurements of $\text{d}t/\text{d}\nu$ for six bursts of roughly increasing duration (left to right), repeated using different methods. These methods are the arrival times pipeline (blue), a 2D Gaussian fit to the burst ACF (orange), and a 2D Gaussian fit directly to the burst waterfall (light and dark gray). The light gray represents a generic 2D Gaussian form (see, e.g., \citealt{Chamma2023}) while the dark gray uses the 2D Gaussian parameterized with physical variables, including the slope introduced by \citet{Jahns2023}. Note that these last two result in identical measurements. The peach points \textbf{(right axis)} show the largest percent difference between the four measurements for each burst. While measurements are almost identical in some cases, there can also be significant differences between them. The bursts shown are from \citet{Snelders2023} (B30 and B43), \citet{Michilli2018} (M001 and M016), and \citet{Li2021} (M01\_0136, M01\_0153) and are all from FRB 20121102A.}
\end{figure}

\begin{figure}
    \includegraphics[width=\columnwidth]{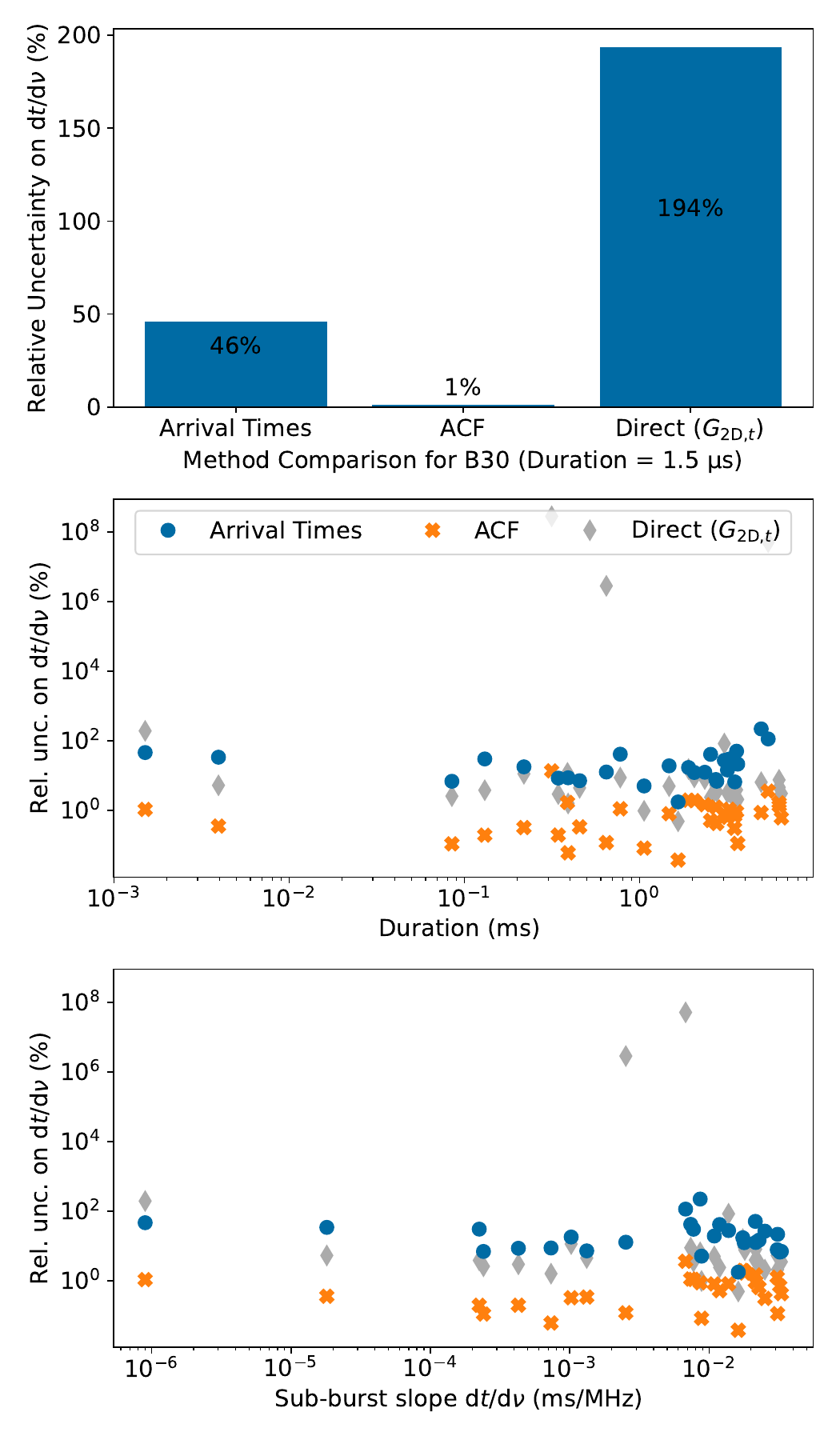}
    \caption{\label{fig:unccompare}Method comparisons for a subsample of bursts, focusing on the resulting uncertainty following each method. \textbf{(top)} Uncertainties between methods focused on the ultra-FRB B30 from \citet{Snelders2023} which has a duration of 1.5 $\upmu$s. \textbf{(center)} Uncertainty for a subsample of bursts as a function of burst duration. The circles, x's and diamonds denote the uncertainty when using the arrival times pipeline, ACF, and direct Gaussian methods, respectively. \textbf{(bottom)} Same as center panel but as a function of the inverse sub-burst slope. Smaller values of d$t$/d$\nu$ indicate a more `vertical' burst. Note that the uncertainties from the arrival times and direct gaussian methods are generally comparable at any scale, while the uncertainties from the ACF method are often more precise.}
\end{figure}

The bursts were selected from FRB 20121102A to cover a broad range of durations, and from datasets with sufficiently high time resolution so that the precision in the arrival times pipeline was not limited. The methods applied to each burst were as follows: (a) the arrival times pipeline, (b) fitting a general 2D Gaussian to the ACF of the burst waterfall, followed by computing measurements using the fit parameters \citep{Chamma2023}, (c) directly fitting a general 2D Gaussian, $G_\text{2D}$, to the waterfall and computing measurements from the fit parameters, and (d) directly fitting a 2D Gaussian with physically defined parameters including the intra-burst drift, $d_t$, to the waterfall \citep[Eq. 2 of][]{Jahns2023}. In total, 34 bursts were analysed, including two ultra-FRBs from \citet{Snelders2023}, nine bursts from \citet{Michilli2018}, and 23 bursts from \citet{Li2021}. This selection was largely arbitrary, as applying multiple measurement methods to each burst is labor-intensive, requiring thorough review and quality assurance for each measurement. Regardless, this subset was sufficient for our purposes. We initially included 29 bursts from \citet{Oostrum2020} in this multi-method analysis, however, the low time resolution of that dataset affected the precision of the arrival time $t_\text{arr}$ measured in each channel (see Eq. \ref{eq:arrtime}), severely limiting the precision of the arrival times method. In such instances, where time resolution is low, the Gaussian methods provide higher precision. Additional bursts were analysed from the datasets listed, but only those yielding quality measurements across all methods were retained. The results of this multi-method comparison are shown in Figures \ref{fig:methcompare} and \ref{fig:unccompare}. 

Figure \ref{fig:methcompare} displays the measurement values obtained for $\text{d}t/\text{d}\nu$ across six bursts of roughly increasing durations. As shown, the measurement values from the four applied methods can align closely, as seen for bursts M001 ($\sigma_t = 0.39$ ms) and M01\_0136 ($\sigma_t = 2.7$ ms). Across these six bursts, the measurement values are generally of the same order of magnitude, even when applied to ultra-FRBs. However, the largest percent difference between values (i.e. the largest difference in measurements divided by the arrival times measurement) can exceed 300\%. These statements generally apply to the entirety of the subset of bursts analysed. Additionally, the uncertainties for each measurement method are shown, and we observe that, for longer-duration bursts (the four right-most bursts; remembering $|\text{d}t/\text{d}\nu| \propto \sigma_t$), uncertainties are small regardless of the method used with the uncertainties from the arrival times method typically being a bit larger. This remains true for the ultra-FRB, B43, however with B30 we see the direct Gaussian methods yield a measurement uncertainty larger than 100\%, the arrival times yields an uncertainty of about 46\%, and the ACF method again yields the best precision with an uncertainty of just 1\%. 

In Figure \ref{fig:unccompare}, the three panels provide additional data highlighting the differences and similarities with the arrival times method. The top panel shows the aforementioned uncertainties for the ultra-FRB B30. The middle and bottom panels qualitatively illustrate the relationship between the uncertainty in $\text{d}t/\text{d}\nu$ and the burst's duration and $\text{d}t/\text{d}\nu$ measurement, respectively. Despite fluctuations, we observe that at all values of duration and sub-burst slope, uncertainties from the arrival times, and direct Gaussian methods are often comparable, and show a flat trend, with uncertainties from the ACF method usually being much smaller. The differences observed arise due to burst waterfall characteristics rather than the method used.  
 
Based on the comparisons shown here, if burst structure and/or blending is not an issue, the ACF method will typically yield higher precision. In waterfalls where burst structure and profiles are complicated, or if blending is present, the arrival times and the direct Gaussian methods may be more appropriate.

\section{Discussion}\label{sec:discussion}
In this section we discuss the physical interpretation of the spectro-temporal relationships observed in our analysis of burst data using the arrival times pipeline. We discuss the effects of interstellar scattering on the observed relationships and on specific burst properties, identifying those most affected by scattering. 

% \subsection{Physical interpretation of spectro-temporal relationships}

% A physical interpretation of the sub-burst slope relation observed in Figure \ref{fig:nudtdnu} is offered by the TRDM of \citet{Rajabi2020}.
% %, along with predictions for other relationships that are not clearly observed in the results presented here. 

A physical interpretation of the sub-burst slope relation observed in Figure \ref{fig:nudtdnu} is offered by the TRDM of \citet{Rajabi2020}. The arrival time $t_\text{arr}$, defined and used here as the basis of our measurement method, is equivalent to the delay time $t_\text{D}$, representing the time interval between the (unobserved) trigger and the start of FRB emission. In the TRDM, the predicted relationship between sub-burst slope and duration is scaled by a physical constant such that, using the notation defined earlier, 
\begin{equation}
    \nu_0\frac{\text{d}t}{\text{d}\nu} = -\Big(\frac{\tau'_\text{D}}{\tau'_\text{w}}\Big)\sigma_t,
\end{equation}
where $\nu_0$ is the observed burst frequency, and our earlier fit parameter $a$ is defined by $a\equiv -(\tau'_\text{D}/\tau'_\text{w})$, which represents the ratio of the proper delay time to the proper burst duration in the rest frame of the FRB source. Thus, within the context of the TRDM, the scaling of the sub-burst slope--duration relation arises from a constant factor linking the delay time following a trigger experienced by an FRB emitter to the resulting duration of an FRB pulse. This model was inspired by the application of Dicke's superradiance to FRBs \citep{Houde2019}, though it is broadly applicable. It is important to note that beyond the assumption of a fundamentally narrowband emission process, the TRDM is agnostic to the underlying emission mechanism and primarily describes the transformation of the FRB signal by dynamical and relativistic effects.

% %This model was proposed within context of Dicke's superradiance \citep{Dicke1954}, posited as the potential underlying physical mechanism. Dicke's superradiance is a cooperative quantum mechanical emission phenomenon characterized by an electromagnetic trigger followed by a delayed response. While signficant work has been done in applying superradiance to astrophysical contexts \citep[e.g.][]{Rajabi2016,Rajabi2017,Houde2018a} and some work in the specific context of FRBs \citep{Houde2017} a detailed analysis taking into account recent observational results on FRBs remains to be done. While the TRDM was inspired in the context of Dicke's superradiance, it is important to note that beyond the assumption of a fundamentally narrowband emission process, the TRDM is agnostic to the underlying emission mechanism and primarily describes the transformation of the FRB signal by dynamical and relativistic effects.

The slope--frequency relation is a strong correlation that is observed in our results here and predicted in the TRDM \citep[Eq. 7 of][]{Chamma2023}. Despite three significant outliers (out of seven total points), the addition of low frequency LOFAR data for FRB 20180916B seems to suggest a related trend between at least that source and FRB 20121102A. This is because the same scaling of the fit shown in the $\text{d}t/\text{d}\nu$--$\nu_0$ panel of Figure \ref{fig:corner} describes most of the FRB 20180916B data and indeed the data from all sources well. In the TRDM the scaling constant for this relationship is equivalent to $1/(\tau'_\text{D}\nu_\text{e})$ where $\nu_\text{e}$ is the rest frame frequency of emission of the FRB. This relationship is also expected in the context of curvature radiation by charged bunches in the magnetosphere of magnetars \citep{Wang2022}. In this vein, a recent scintillation analysis for FRB 20221022A argues that the lateral size of the FRB emission region in that source to be less than $\sim3\times10^4$ km \citep{Nimmo2025}.

The sub-burst slope relation in the TRDM is derived directly from a more basic relationship between burst duration and frequency, and, with some caveats, there is good evidence of this basic relation in our results. Namely, the TRDM expects $\sigma_t \propto \nu_0^{-1}$ \citep[Eqs. 3 and 4 of][note the differing definition of $\nu_0$]{Rajabi2020}. As discussed in Section \ref{subsubsec:relations}, the addition of ultra-FRBs in this analysis makes it clear that while the ultra-FRBs appear to obey this relation, they can form a distinct group on the $\sigma_t$--$\nu_0$ plot \citep{Kumar2024}. Other forms might also describe this data, given the spread in measurements.

% A natural question arises in the context of the TRDM; how can the sub-burst slope relation appear to be valid while at least one associated underlying relationship ($\sigma_t \propto \nu_0^{-1}$) lacks clear evidence? This on one hand can be interpreted as a challenge to the TRDM's assumptions.

A likely scenario is that the underlying TRDM relations do hold for FRBs but are obfuscated by physical processes, either before, during, or after emission, that add statistical scatter to the observed burst properties. This statistical scatter spans several orders of magnitude for bursts analysed here (such as in the two orders of magnitude spanned by duration measurements for bursts around 1500 MHz) and is an important reason that most burst parameters appear uncorrelated in many narrowband studies. It is therefore important to characterize the extent of this statistical scatter since, assuming the validity of the TRDM's simple assumptions, they may arise directly from the underlying emission mechanism. 

An explanation for the behavior of ultra-FRBs in the sub-burst slope law (Figure \ref{fig:nudtdnu} and inset) and $\sigma_t$--$\nu_0$ relation is put forward by \citet{Kumar2024}. They show through simulation that the effects of interstellar scattering and residual dispersion modifies the sub-burst slope law in a way that affects ultra-FRBs differently from longer duration FRBs. This includes the presence of a positive sub-burst slope `bump' for ultra-FRBs, which we can observe in our results.

% \subsection{Effects of interstellar scattering}

Interstellar scattering can significantly affect the durations of measured bursts and, in principle, delay the arrival times of pulses in a frequency-dependent manner, thereby impacting the measurement of the sub-burst slope. The extent of this effect depends on the scattering timescale, which quantifies the amount of scattering present. For sources with significant scattering, deviations from the ideal sub-burst slope law can be substantial, particularly at low frequencies and for intrinsically short duration bursts \citep{Kumar2024}. At shorter scattering timescales, these deviations become less pronounced. Since the ultra-FRBs used here were at a higher frequency, the impact of scattering on this study is likely modest.

In particular, the inverse of the sub-burst slope of ultra-FRBs can be shifted upwards (more positive) relative to that of standard FRBs when plotted against their duration. This occurs even for scattering timescales approximately as short as their intrinsic (i.e., without scattering) duration \citep[][their fig. 9]{Kumar2024}. We note however that \citet{Nimmo2023} report an absence of scattering in their data. \citet{Sheikh2024} also do not see significant scattering, but cannot rule out scattering timescales below 0.06 ms. For FRB 20121102A, \citet{Michilli2018} also report no significant scattering and constrain the scatter broadening to around 0.024 ms at 1 GHz. In contrast, the LOFAR points for FRB 20180916B, which are of much lower frequency, do exhibit long scattering timescales, on the order of 50 ms \citep{Pleunis2021}. Of these, 3 of 7 points appear as outliers in Figures \ref{fig:nudtdnu} and \ref{fig:corner}, and have slopes with smaller magnitude than expected for the d$t$/d$\nu-\sigma_t$ and d$t$/d$\nu-\nu_0$ relations, consistent with predictions by \citet{Kumar2024}. There are therefore potentially observable effects on our results due to scattering. If so, it is most likely a limited effect for most of our results but could explain minor deviations from the d$t$/d$\nu-\sigma_t$, and other, expected relations.

We also note that the sub-burst slope of ultra-FRBs is especially sensitive to dispersion errors. Because of their extremely short duration, ultra-FRBs appear close to vertical on a perfectly dedispersed waterfall. It follows that even small levels of over-dedispersion can change the sign of their sub-burst slope \citep[][their fig. 10]{Kumar2024}. In principle, this could explain the tendency for some of the sub-burst slopes of shorter duration bursts, especially those from FRB 20200120E (durations between $10^{-2}$ and $10^{-1}$ ms) and FRB 20121102A and FRB 20220912A (durations above $10^{-1}$ ms), to `bump' upwards in Figure \ref{fig:nudtdnu}.

One natural extension of the arrival times method is measuring and quantifying interstellar scattering on FRBs. This can be done by replacing the Gaussian profile fit to each frequency channel of an FRB waterfall (Eq. \ref{eq:profile}) with a profile shape that consists of a Gaussian convolved with a scattering tail. 

This can serve as a more accurate model for FRBs that are less Gaussian-like in their pulse shape while also providing an optimized measurement of the scattering timescale as a function of frequency. %It may further be advantageous in increasing the accuracy of the 1D model of the burst time series, especially for complicated waterfalls.

Further adjustments would be necessary to integrate this change into the pipeline, such as updating the definition of duration used, and some care is needed to see how the pipeline behaves for both bursts that exhibit scattering and those that do not. We leave this analysis to future research efforts. We also note that while bursts with potential scattering tails were observed in our sample, they were the vast minority and the majority were well described by a Gaussian profile.

Evidently, there are rich connections between observed spectro-temporal relationships and the physical models surrounding FRBs. We therefore believe that continued and deep monitoring of repeating FRB sources with the intent of matching the diversity of bursts available for FRB 20121102A would be invaluable to this type of spectro-temporal analysis and provide significant insights and constraints on FRB emission mechanisms.

\section{Summary and Conclusions}\label{sec:conclusion}

We have developed and applied a measurement pipeline that uses the arrival times of an FRB pulse in each frequency channel of its dynamic spectrum to perform sub-burst slope/intra-burst drift measurements. The pipeline successfully delivered high precision measurements of the slope including for ultra-FRBs. In most cases, earlier 2D Gaussian techniques yielded measurements with similar values and uncertainties, whereas the ACF method yielded the smallest uncertainties across all methods used. Drift rates were also measured for bursts with multiple resolved components, yielding a total of 143 measurements.

The arrival times pipeline was successfully applied to a total of 433 FRBs from 12 repeating sources. The focus of our analysis was on the repeating sources FRB 20121102A, FRB 20220912A, and FRB 20200120E, all of which have recently exhibited ultra-FRBs. Bursts from FRB 20121102A spanned the broadest duration and frequency range in our sample, largely due to the many years of observation this source has enjoyed. These bursts along with the remaining data span 110 MHz up to 8 GHz in frequency and 1.6 $\upmu$s up to 4.2 ms in duration, and therefore constitute a sample of bursts that cover several orders of magnitude of burst properties as well as morphologies. Sub-burst slope $\text{d}t/\text{d}\nu$, center frequency $\nu_0$, bandwidth $\sigma_\nu$, and duration $\sigma_t$ measurements were obtained for each of these bursts using the arrival times pipeline.

Analysing the spectro-temporal relationships between these properties showed once again the known strong correlation between sub-burst slope and duration, the scaling of which was consistent with that found in earlier studies using the 2D Gaussian techniques. For bursts from FRB 20121102A, we found $\nu(\text{d}t/\text{d}\nu)\simeq (-8.6 \pm 0.4) \sigma_{t}$. Other strong correlations were observed between the $\text{d}t/\text{d}\nu$--$\sigma_\nu$ and $\text{d}t/\text{d}\nu$--$\nu_0$ relations, the latter showing a single scaling describing the data from all sources well. We see a correlation between $\sigma_t$ and $\nu_0$, though the ultra-FRBs appear distinct from the longer duration FRBs. For the remaining $\sigma_t$--$\sigma_\nu$ and $\nu_0$--$\sigma_\nu$ relations, there was significant statistical scatter in measurements. For example, at $\sim$1500 MHz, the measured durations span nearly four orders of magnitude. The drift rates measured showed good agreement with the analogous relation between sub-burst slope and duration and seem to extend the trend laid by the sub-burst slopes.

Comparing measurements obtained from the arrival times pipeline with multiple 2D Gaussian techniques, including fitting to the burst ACF and fitting directly to the burst waterfall, generally revealed agreement in measurement values between the methods with certain factors to keep in mind that could lead to significant differences. Differences are generally attributable to complicated burst morphologies or RFI. In that regard, measurements from the arrival times pipeline can be quite sensitive to RFI or a bad channel, since a few misplaced arrival times can skew a measurement. The arrival times pipeline can also be limited and give higher relative uncertainties than Gaussian methods if the time resolution of the data is insufficiently high. 

A simple linear model was applied when obtaining measurements of the sub-burst slope $\text{d}t/\text{d}\nu$ in this study, however the arrival times method offers straightforward extensions that can accommodate more complicated drifting morphologies. 

While interstellar scattering and the residual DM of a source remain significant and often dominating factors in the accuracy (or inaccuracy) of spectro-temporal measurements, the arrival times pipeline offers an adaptable and firm foundation for obtaining numerically precise spectro-temporal measurements from FRBs. The arrival times method is available in an online package with documentation and open to contributions from the community.

\section*{Acknowledgements}
The authors would like to thank the anonymous referees, whose feedback greatly improved the quality of the manuscript and its conclusions.

We would also like to thank Sofia Sheikh, Wael Farah, Danté Hewitt, Mark Snelders, and Yong-Kun Zhang for their help in accessing their respective datasets and for additional assistance in reading it. We are also grateful to Martin Houde for their extensive feedback and discussions of the text.

F.R.'s research is supported by the Natural Sciences and Engineering Research Council of Canada (NSERC) Discovery Grant RGPIN-2024-06346.

%%%%%%%%%%%%%%%%%%%%%%%%%%%%%%%%%%%%%%%%%%%%%%%%%%
\section*{Data Availability}

% The inclusion of a Data Availability Statement is a requirement for articles published in MNRAS. Data Availability Statements provide a standardised format for readers to understand the availability of data underlying the research results described in the article. The statement may refer to original data generated in the course of the study or to third-party data analysed in the article. The statement should describe and provide means of access, where possible, by linking to the data or providing the required accession numbers for the relevant databases or DOIs.

All measurements and scripts produced during the analysis are available online at \url{https://zenodo.org/records/13357030}.

The arrival times pipeline is packaged with FRBGUI, available at \url{https://github.com/mef51/frbgui}. Technical documentation and tutorials on using the arrival times pipeline is available at \href{https://frbgui.readthedocs.io/en/latest/documentation/doc-arrivaltimes.html}{\texttt{https://frbgui.readthedocs.io/arrivaltimes}}.



\bibliographystyle{mnras}
\bibliography{frblibrary} 
%%%%%%%%%%%%%%%%%%%%%%%%%%%%%%%%%%%%%%%%%%%%%%%%%%

%%%%%%%%%%%%%%%%% APPENDICES %%%%%%%%%%%%%%%%%%%%%

\appendix

\section{Table of ultra-FRB measurements}\label{app:ultrafrbs}

We provide an abridged table of measurements obtained for a subset of ultra-FRBs in Table \ref{tab:appultra}. The complete table including measurements of all bursts and additional columns is available at \url{https://zenodo.org/records/13357030}.

\begin{figure*}
    \centering
    \includegraphics[width=0.9\textwidth]{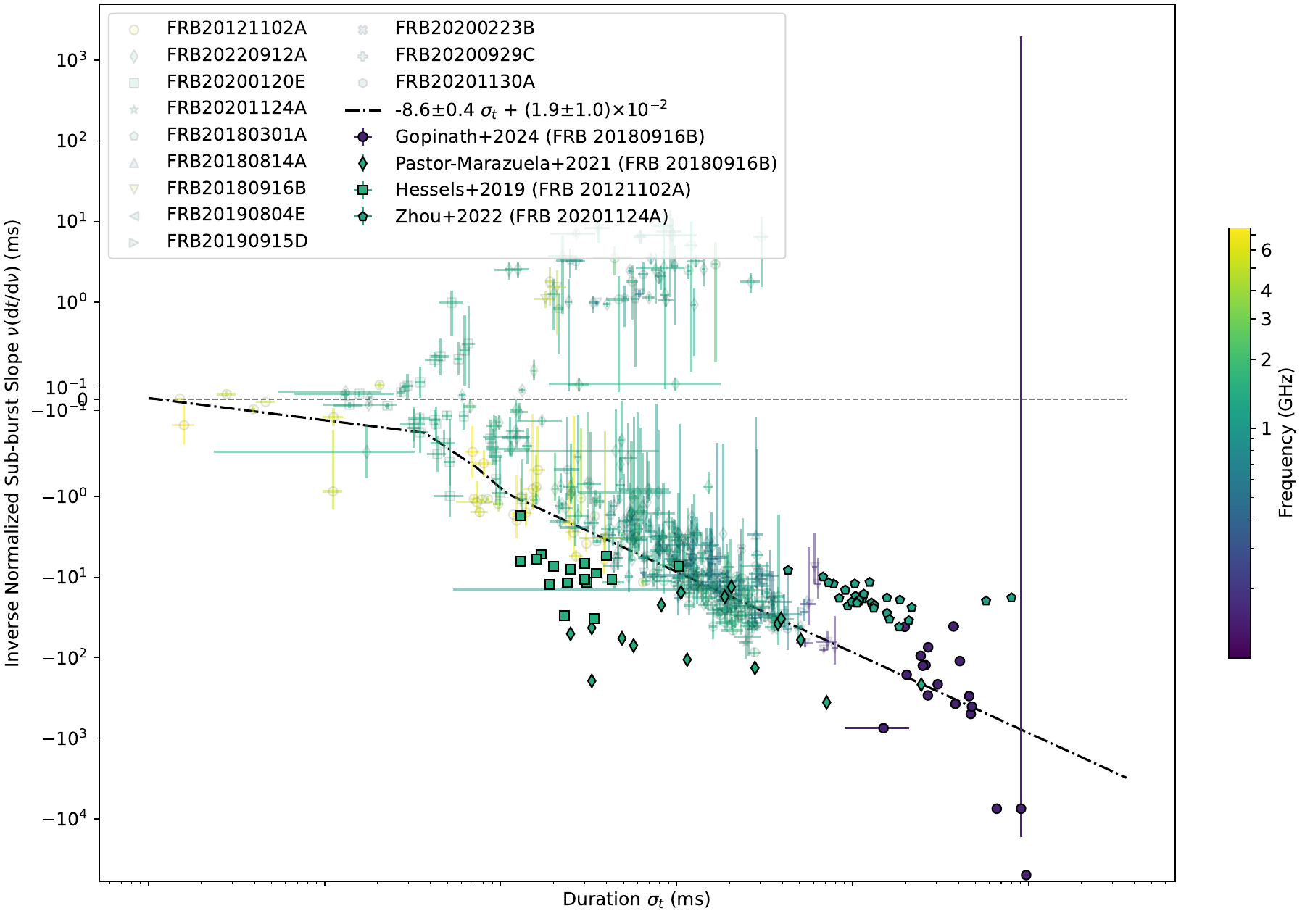}
    \caption{Figure \ref{fig:nudtdnu} but with additional measurements from other studies, with previously shown points faded out and multi-component drift rates removed to aid with clarity. Note also the symmetric-log scale on the y-axis. The additional data shown are from \citet{Gopinath2024} and \citet{PastorMarazuela2021} from FRB 20180916B, \citet{Hessels2019} from FRB 20121102A, and \citet{Zhou2022} from FRB 20201124A.}
    \label{fig:extradata}
\end{figure*}

Additionally, Figure \ref{fig:extradata} shows the sub-burst slope relation plot with slope/drift measurements from additional studies.

\begin{table*}
\begin{centering}
\begin{tabular}{lllrrrr}
\toprule
\textbf{Source} & \textbf{Burst ID} &\textbf{DM pc/cm$^3$} & \textbf{$\nu_0$ MHz} & \textbf{$\sigma_t$ $\upmu$s} & \textbf{$\sigma_\nu$ MHz} & \textbf{$\text{d}t/\text{d}\nu$ ms/MHz} \tabularnewline
\midrule
\midrule
FRB 20121102A & burst-B30 & 560.105 & 6053 $\pm$ 25 & 1.51 $\pm$ 0.1& 324 $\pm$ 26 & (9.0$\pm$4.1)$\times 10^{-7}$ \tabularnewline
 & burst-B06-b & 560.105 & 7075 $\pm$ 12 & 1.59 $\pm$ 0.2& 68 $\pm$ 13 & (-3.3$\pm$2.5)$\times 10^{-5}$ \tabularnewline
 & burst-B07 & 560.105 & 4759 $\pm$ 15 & 2.78 $\pm$ 0.3& 73 $\pm$ 16 & (9.5$\pm$5.4)$\times 10^{-6}$ \tabularnewline
 & burst-B43 & 560.105 & 4928 $\pm$ 10 & 3.95 $\pm$ 0.1& 112 $\pm$ 10 & (-1.8$\pm$0.6)$\times 10^{-5}$ \tabularnewline
 & burst-B10 & 560.105 & 5244 $\pm$ 52 & 4.63 $\pm$ 0.6& 246 $\pm$ 57 & (-4.6$\pm$0.3)$\times 10^{-6}$ \tabularnewline
 & burst-B31-b & 560.105 & 5117 $\pm$ 11 & 11.17 $\pm$ 1.4& 48 $\pm$ 12 & (-1.7$\pm$1.1)$\times 10^{-4}$ \tabularnewline
 & burst-B38 & 560.105 & 5551 $\pm$ 52 & 11.24 $\pm$ 1.5& 330 $\pm$ 56 & (-2.9$\pm$1.5)$\times 10^{-5}$ \tabularnewline
 & M006 & 560.105 & 4447 $\pm$ 16 & 20.55 $\pm$ 1.3& 208 $\pm$ 20 & (2.9$\pm$0.6)$\times 10^{-5}$ \tabularnewline
 & 11A-b & 560.105 & 6984 $\pm$ 30 & 69.33 $\pm$ 6.1& 377 $\pm$ 32 & (-6.7$\pm$3.3)$\times 10^{-5}$ \tabularnewline
 & 11Q & 560.105 & 6054 $\pm$ 16 & 70.46 $\pm$ 2.9& 289 $\pm$ 16 & (-1.8$\pm$0.2)$\times 10^{-4}$ \tabularnewline
 & 11G & 560.105 & 5752 $\pm$ 28 & 73.26 $\pm$ 17.2& 182 $\pm$ 30 & (-2.0$\pm$0.8)$\times 10^{-4}$ \tabularnewline
 & 11N & 560.105 & 5690 $\pm$ 15 & 76.12 $\pm$ 8.7& 169 $\pm$ 15 & (-2.8$\pm$0.4)$\times 10^{-4}$ \tabularnewline
 & M013-b & 560.105 & 4729 $\pm$ 36 & 78.22 $\pm$ 2.8& 163 $\pm$ 33 & (-2.3$\pm$0.3)$\times 10^{-4}$ \tabularnewline
 & 11A-c & 560.105 & 6340 $\pm$ 18 & 80.17 $\pm$ 10.7& 244 $\pm$ 18 & (-9.0$\pm$1.9)$\times 10^{-5}$ \tabularnewline
 & M014 & 560.105 & 4443 $\pm$ 19 & 84.75 $\pm$ 1.3& 223 $\pm$ 25 & (-2.4$\pm$0.2)$\times 10^{-4}$ \tabularnewline
 & M013-c & 560.105 & 4650 $\pm$ 22 & 97.04 $\pm$ 6.3& 165 $\pm$ 25 & (-2.6$\pm$0.6)$\times 10^{-4}$ \tabularnewline
 & 11H & 560.105 & 7262 $\pm$ 11 & 118.33 $\pm$ 2.4& 277 $\pm$ 11 & (-2.3$\pm$0.1)$\times 10^{-4}$ \tabularnewline
 & \textbf{...} & \textbf{...} & \textbf{...} & \textbf{...} & \textbf{...} & \textbf{...} \tabularnewline
\midrule
FRB 20220912A & B1-h & 219.356 & 1533 $\pm$ 88 & 13.17 $\pm$ 7.7& 57 $\pm$ 88 & (4.3$\pm$1.6)$\times 10^{-5}$ \tabularnewline
 & B1-k & 219.356 & 1530 $\pm$ 57 & 17.38 $\pm$ 15.0& 61 $\pm$ 57 & (-3.1$\pm$1.5)$\times 10^{-4}$ \tabularnewline
 & B1-o & 219.356 & 1519 $\pm$ 43 & 17.83 $\pm$ 8.0& 62 $\pm$ 43 & (-3.1$\pm$1.3)$\times 10^{-5}$ \tabularnewline
 & B2-i & 219.375 & 1436 $\pm$ 9 & 18.03 $\pm$ 0.6& 85 $\pm$ 9 & (9.7$\pm$5.2)$\times 10^{-6}$ \tabularnewline
 & 9-06-g & 219.356 & 1227 $\pm$ 3 & 60.48 $\pm$ 2.7& 106 $\pm$ 3 & (2.6$\pm$1.9)$\times 10^{-5}$ \tabularnewline
 & B24-a & 219.356 & 1979 $\pm$ 4 & 67.36 $\pm$ 5.7& 137 $\pm$ 4 & (-3.3$\pm$3.0)$\times 10^{-5}$ \tabularnewline
 & B01-b & 219.356 & 1598 $\pm$ 5 & 92.09 $\pm$ 9.3& 103 $\pm$ 6 & (-1.5$\pm$0.6)$\times 10^{-4}$ \tabularnewline
 & B12-d & 219.356 & 1619 $\pm$ 8 & 98.36 $\pm$ 2.3& 163 $\pm$ 8 & (-1.3$\pm$0.4)$\times 10^{-4}$ \tabularnewline
 & B2-r & 219.375 & 1387 $\pm$ 3 & 111.82 $\pm$ 19.4& 74 $\pm$ 3 & (1.8$\pm$0.4)$\times 10^{-3}$ \tabularnewline
 & B14-c & 219.356 & 1329 $\pm$ 4 & 113.20 $\pm$ 33.6& 82 $\pm$ 4 & (-2.5$\pm$1.3)$\times 10^{-4}$ \tabularnewline
 & B1-t & 219.356 & 1541 $\pm$ 157 & 114.83 $\pm$ 32.7& 62 $\pm$ 157 & (-2.9$\pm$0.4)$\times 10^{-4}$ \tabularnewline
 & B2-q & 219.375 & 1406 $\pm$ 5 & 122.13 $\pm$ 14.4& 79 $\pm$ 5 & (-2.0$\pm$0.6)$\times 10^{-4}$ \tabularnewline
 & B01-d & 219.356 & 1539 $\pm$ 5 & 122.58 $\pm$ 9.7& 104 $\pm$ 5 & (-6.1$\pm$5.9)$\times 10^{-5}$ \tabularnewline
 & 9-06-l & 219.356 & 1164 $\pm$ 1 & 123.72 $\pm$ 21.6& 58 $\pm$ 1 & (-3.0$\pm$0.8)$\times 10^{-4}$ \tabularnewline
 & B04-b & 219.356 & 1357 $\pm$ 5 & 125.50 $\pm$ 20.2& 52 $\pm$ 5 & (1.9$\pm$0.4)$\times 10^{-3}$ \tabularnewline
 & B01-c & 219.356 & 1565 $\pm$ 6 & 127.48 $\pm$ 15.3& 102 $\pm$ 7 & (-7.2$\pm$5.1)$\times 10^{-5}$ \tabularnewline
 & \textbf{...} & \textbf{...} & \textbf{...} & \textbf{...} & \textbf{...} & \textbf{...} \tabularnewline
\midrule
FRB 20200120E & 20220223-B1 & 87.7527 & 1385 $\pm$ 0 & 13.10 $\pm$ 0.7& 8 $\pm$ 0 & (2.4$\pm$2.2)$\times 10^{-5}$ \tabularnewline
 & B3-a & 87.75 & 1346 $\pm$ 4 & 13.86 $\pm$ 2.6& 46 $\pm$ 5 & (-3.9$\pm$1.7)$\times 10^{-5}$ \tabularnewline
 & B3-b & 87.75 & 1331 $\pm$ 4 & 15.78 $\pm$ 9.0& 37 $\pm$ 4 & (3.6$\pm$2.2)$\times 10^{-5}$ \tabularnewline
 & B2-a & 87.75 & 1344 $\pm$ 5 & 22.75 $\pm$ 0.7& 41 $\pm$ 5 & (-4.4$\pm$2.2)$\times 10^{-5}$ \tabularnewline
 & 20220114-B47 & 87.7527 & 1326 $\pm$ 7 & 27.13 $\pm$ 1.5& 73 $\pm$ 8 & (4.8$\pm$3.2)$\times 10^{-5}$ \tabularnewline
 & 20220114-B37 & 87.7527 & 1430 $\pm$ 11 & 28.59 $\pm$ 1.4& 96 $\pm$ 13 & (7.6$\pm$2.9)$\times 10^{-5}$ \tabularnewline
 & 20220114-B49 & 87.7527 & 1398 $\pm$ 11 & 29.59 $\pm$ 2.1& 85 $\pm$ 12 & (8.5$\pm$7.9)$\times 10^{-5}$ \tabularnewline
 & 20220114-B32 & 87.7527 & 1403 $\pm$ 15 & 32.05 $\pm$ 5.0& 83 $\pm$ 16 & (-1.6$\pm$1.1)$\times 10^{-4}$ \tabularnewline
 & 20220114-B34 & 87.7527 & 1443 $\pm$ 8 & 32.18 $\pm$ 2.5& 79 $\pm$ 9 & (-1.8$\pm$1.2)$\times 10^{-4}$ \tabularnewline
 & 20220114-B42 & 87.7527 & 1408 $\pm$ 13 & 34.81 $\pm$ 4.8& 66 $\pm$ 13 & (-1.2$\pm$0.2)$\times 10^{-4}$ \tabularnewline
 & 20220114-B18-b & 87.7527 & 1361 $\pm$ 12 & 34.83 $\pm$ 4.2& 83 $\pm$ 13 & (-2.2$\pm$1.3)$\times 10^{-4}$ \tabularnewline
 & 20220114-B50 & 87.7527 & 1421 $\pm$ 16 & 34.88 $\pm$ 2.3& 80 $\pm$ 16 & (1.1$\pm$1.0)$\times 10^{-4}$ \tabularnewline
 & 20220114-B48 & 87.7527 & 1365 $\pm$ 17 & 42.14 $\pm$ 4.9& 96 $\pm$ 18 & (2.6$\pm$0.5)$\times 10^{-4}$ \tabularnewline
 & 20220114-B13-b & 87.7527 & 1358 $\pm$ 14 & 42.32 $\pm$ 3.1& 94 $\pm$ 16 & (-1.4$\pm$0.6)$\times 10^{-4}$ \tabularnewline
 & B4 & 87.75 & 1382 $\pm$ 5 & 43.06 $\pm$ 1.1& 55 $\pm$ 5 & (-1.6$\pm$1.1)$\times 10^{-4}$ \tabularnewline
 & 20220114-B23 & 87.7527 & 1374 $\pm$ 16 & 43.72 $\pm$ 5.3& 93 $\pm$ 17 & (-3.6$\pm$1.2)$\times 10^{-4}$ \tabularnewline
 & 20220114-B13-a & 87.7527 & 1363 $\pm$ 24 & 45.67 $\pm$ 5.5& 116 $\pm$ 28 & (2.8$\pm$1.2)$\times 10^{-4}$ \tabularnewline
 & 20220114-B22 & 87.7527 & 1390 $\pm$ 17 & 47.57 $\pm$ 7.6& 88 $\pm$ 18 & (-2.8$\pm$0.9)$\times 10^{-4}$ \tabularnewline
 & 20220114-B21 & 87.7527 & 1415 $\pm$ 13 & 49.56 $\pm$ 3.4& 72 $\pm$ 13 & (-1.0$\pm$0.3)$\times 10^{-4}$ \tabularnewline
 & 20220114-B26 & 87.7527 & 1358 $\pm$ 13 & 51.36 $\pm$ 3.8& 84 $\pm$ 13 & (-4.1$\pm$1.6)$\times 10^{-4}$ \tabularnewline
 & 20220114-B29 & 87.7527 & 1296 $\pm$ 4 & 51.63 $\pm$ 9.9& 28 $\pm$ 4 & (-7.6$\pm$6.2)$\times 10^{-4}$ \tabularnewline
 & 20220114-B53 & 87.7527 & 1455 $\pm$ 8 & 52.75 $\pm$ 8.2& 41 $\pm$ 8 & (6.7$\pm$2.9)$\times 10^{-4}$ \tabularnewline
 & 20220114-B31 & 87.7527 & 1415 $\pm$ 9 & 57.72 $\pm$ 2.8& 75 $\pm$ 10 & (2.5$\pm$1.1)$\times 10^{-4}$ \tabularnewline
 & 20220114-B35 & 87.7527 & 1398 $\pm$ 17 & 61.88 $\pm$ 3.8& 85 $\pm$ 18 & (-1.1$\pm$0.9)$\times 10^{-4}$ \tabularnewline
 & 20220114-B20 & 87.7527 & 1393 $\pm$ 16 & 62.60 $\pm$ 4.1& 104 $\pm$ 17 & (3.1$\pm$2.3)$\times 10^{-4}$ \tabularnewline
 & \textbf{...} & \textbf{...} & \textbf{...} & \textbf{...} & \textbf{...} & \textbf{...} \tabularnewline
\midrule
\bottomrule
\end{tabular}
\par\end{centering}
\caption{Abridged table of spectro-temporal measurements obtained for a subset of ultra-FRBs using the arrival times method. Rows are organized by source and sorted by the sub-burst duration $\sigma_t$ in ascending order. The full table with measurements of all 433 bursts and additional columns is available online at \url{https://zenodo.org/records/13357030}.}\label{tab:appultra}
\end{table*}

%  Paste caption in:
% Abridged table of spectro-temporal measurements obtained for a subset of ultra-FRBs using the arrival times method. Rows are organized by source and sorted by the sub-burst duration $\sigma_t$ in ascending order. The full table with measurements of all XXX bursts and additional columns is available online at \url{https://zenodo.org/records/13357030}.

\section{Measuring microshot forests with the arrival times pipeline}\label{app:microshot}

\begin{figure*}
    \includegraphics[width=\textwidth]{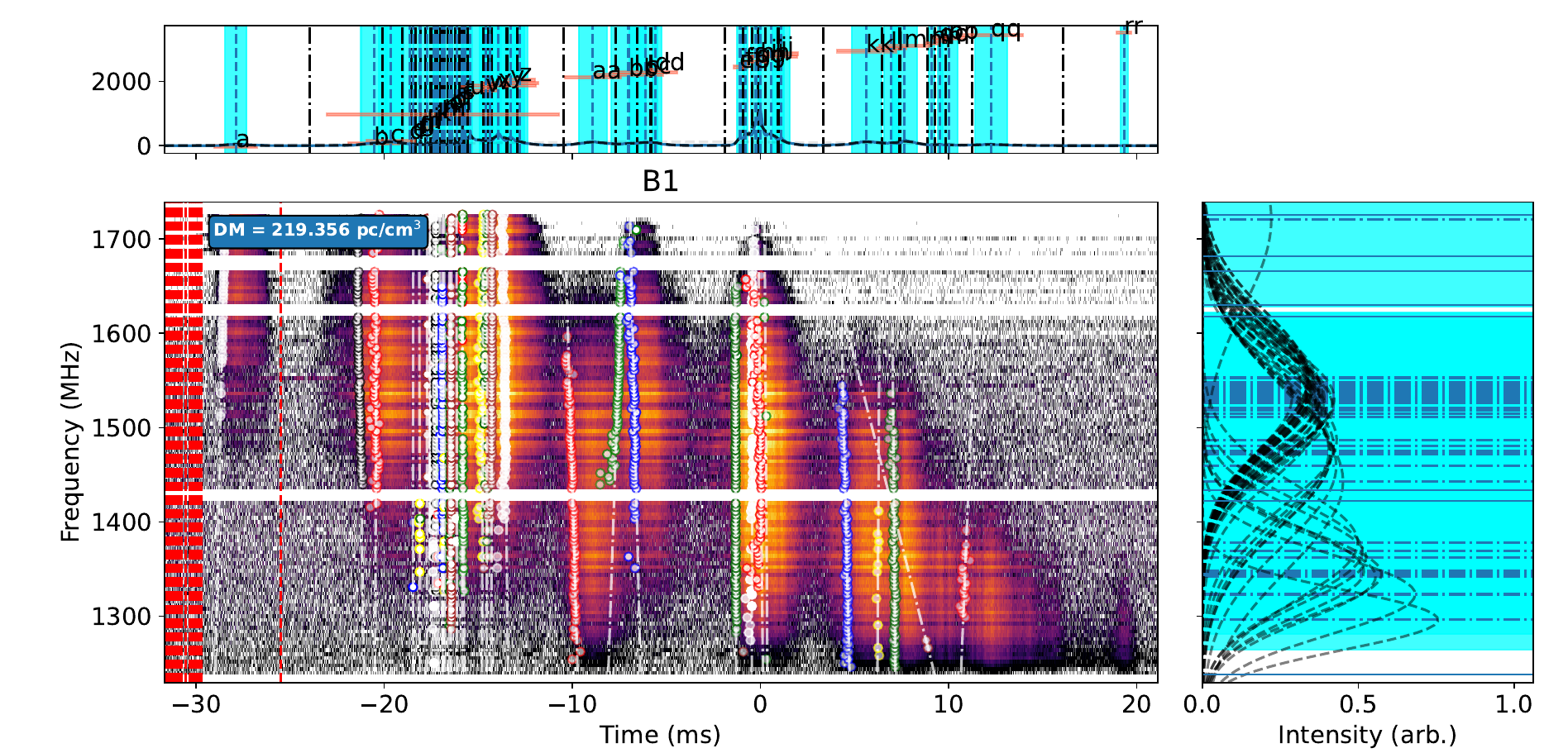}
    \includegraphics[width=\textwidth]{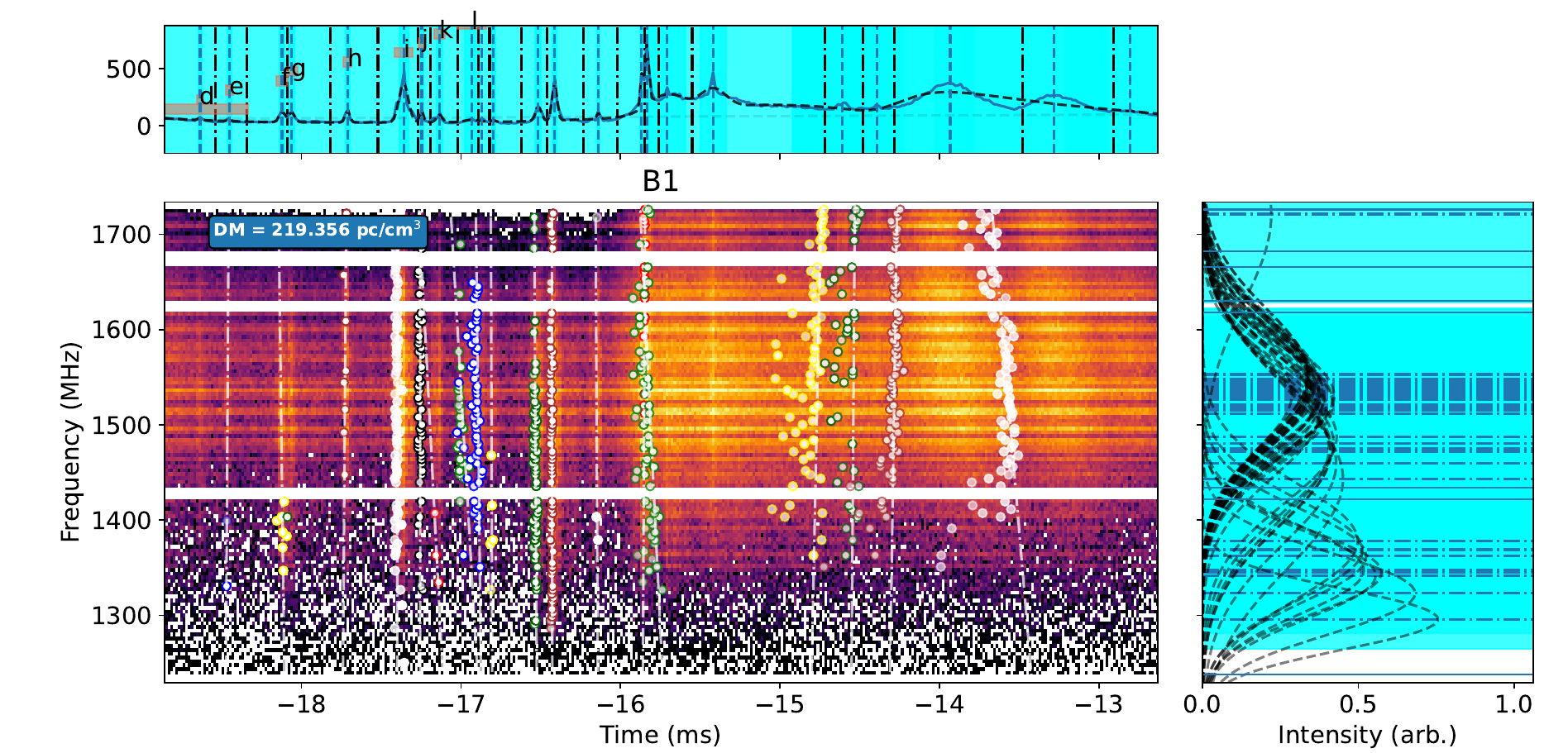}
    \caption{\label{fig:b1}Results of analysing microshot forest burst B1 of \citet{Hewitt2023} with the arrival times pipeline using a manual lists of burst positions (blue dashed lines) and widths as initial guesses and manual cuts (black dash-dot lines) to separate components. Top waterfall shows the full B1 burst, while the bottom shows a zoomed in view of a very active few milliseconds starting at around -18 ms with many sub-millisecond pulses. This bottom panel demonstrates well the challenges of analysing microshots; we see several pulses are well measured, several more have inaccurate sub-burst slope measurements, and a few still are missed completely due to their blended state, short duration, and/or faintness.}
\end{figure*}

Though challenging, the arrival times pipeline can be used to obtain measurements for the components of bursts that exhibit dense microshot forests with dozens of components, including the examples analysed here from \citet{Hewitt2023} and \citet{Zhang2023}. We briefly describe our approach to obtaining measurements from these waterfalls as an example of applying the arrival times pipeline to such bursts. We also discuss the challenges and issues faced, including components that could not be usefully measured due to significant blending.

As explained in Section \ref{sec:methods}, an early step of the arrival times pipeline is to find a fit to the one dimensional integrated time series of the waterfall. For burst B1 of \citet{Hewitt2023}, we note 44 components can be distinguished visually from the time series. The large number of components poses a significant computational challenge to the fitting algorithm. Thus it is important as much as possible to provide a strong initial guess to the algorithm for there to be an accurate solution found in a reasonable amount of time. In the arrival times pipeline, this means providing a long list of $t_0$s specifying the precise time of each of the 44 burst components. Additionally, we specified a corresponding list of pulse durations $\sigma_t$s that was necessary for a reasonably accurate solution. Because of the prevalent presence of blending, we also carefully chose the locations of manual cuts between components as best as possible to indicate what data should be cutout when finding fits in each frequency channel. Despite the success of this in several components, if two components are significantly blended together then the manual cut, other than being difficult to place, is also insufficient in overcoming the fact that the data of the two blended components obscure one another. With a good initial guess it takes about 20 minutes of computation on an M2 chip on a laptop to perform the measurements of burst B1's components. 

The result of applying this process yields many usable measurements but also many unusable measurements, either because of large uncertainties caused by blending or the difficulty of visually confirming the measurement of a faint or blended component. For example, for burst B1 \citet{Hewitt2023}, of the 44 components identified manually, only about 20 are used, and the rest are excluded due to large relative uncertainties on their duration or sub-burst slope greater than 100\%, an insufficient number of well fitted channels to obtain a complete measurement, or some combination of these three issues. In particular, we were unable to obtain measurements for multiple of the narrow components because they were blended with a broader component. This can prevent an appropriately narrow profile from being found in each frequency channel (even when the S/N is high) and the arrival times subsequently found fail to pass the requirement of being within 2$\sigma$ of the component time (outlined in step (v) of Section \ref{subsec:arrtime}).

Figure \ref{fig:b1} shows results from the arrival times pipeline for burst B1 that displays the range of outcomes found in measuring the microshots. Many components are measured well, whereas others are missed or yield measurements with large uncertainties.

\section{DM optimization using the sub-burst slope relation}\label{subsec:dmopt}

\begin{table}
\begin{centering}
\begin{tabular}{llcccc}
\toprule 
\multicolumn{6}{c}{\textbf{FRB 20121102A}}\tabularnewline
\midrule 
\textbf{Step} & \textbf{$\text{\textbf{DM}}_{\text{app}}$} & $b$ (MHz$^{-1}$) & $c$ (ms/MHz) & $\delta$DM & $\text{DM}_{\text{real}}$\tabularnewline
\midrule
\midrule 
\textbf{1.} & 560.105 & $-3.8\times10^{-3}$ & $8.0\times10^{-6}$ & $-0.023$ & 560.082\tabularnewline
\midrule 
\textbf{2.} & 560.082 & $-3.94\times10^{-3}$ & $2.9\times10^{-5}$ & $-0.093$ & 559.989\tabularnewline
\midrule 
\textbf{3.} & 559.989 & $-4.1\times10^{-3}$ & $9.1\times10^{-6}$ & $-0.024$ & 559.965\tabularnewline
\midrule 
\midrule
\textbf{1b.} & 562.41 & $-2.9\times10^{-3}$ & $2.0\times10^{-4}$ & $-0.438$ & 561.972\tabularnewline
\midrule
\textbf{2b.} & 561.927 & $-2.9\times10^{-3}$ & $1.86\times10^{-4}$ & $-0.435$ & 561.537\tabularnewline
\midrule
\textbf{3b.} & 561.537 & $-3.5\times10^{-3}$ & $2.3\times10^{-4}$ & $-0.528$ & 561.010\tabularnewline
\midrule
\multicolumn{6}{c}{\textbf{FRB 20220912A}}\tabularnewline
\midrule 
\textbf{1.} & 219.356 & $-6.55\times10^{-3}$ & $8.98\times10^{-4}$ & $-0.208$ & 219.149\tabularnewline
\midrule 
\textbf{2.} & 219.149 & $-7.1\times10^{-3}$ & $1.3\times10^{-3}$ & $-0.302$ & 218.847\tabularnewline
\bottomrule
\end{tabular}
\par\end{centering}
\caption{DM Bootstrapping for FRB 20121102A and FRB 20220912A. Note that we
exclude the microshot bursts for FRB 20220912A from \citet{Hewitt2023} due to significant blending. Units of DM are pc/cm$^3$. Step 1b indicates a different starting DM.}\label{tab:dm}
\end{table}

Assuming the sub-burst slope relation applies to all FRBs, it can be used to define an optimal or corrected DM that is associated with the repeating FRB source. We summarize here strategies from the literature that have been used so far to do so and describe our own efforts with the measurements presented here.

Given a set of measurements of spectro-temporal properties from FRBs one must decide what  DM at which to display them. A simple choice is to display the measurements at each burst's individually determined DM. This is a valid choice and can be sufficient. However, given the expected relationship between the sub-burst slope and duration, we can apply a single DM to a cohort of bursts (preferably bursts that were emitted close in time to account for long-term changes in the DM) and impose the assumption that bursts that deviate from the sub-burst slope relation are over-- or under-- dedispersed. Because the measurements of $\text{d}t/\text{d}\nu$ are affected by the DM applied, this implies the existence of an optimal DM where this assumption holds best. We may interpret this optimal DM as the `real' DM, absent of contributions to the DM that arise from propagation away from the source. That is, assuming the veracity of the sub-burst slope relation, it can be used to correct the observed DM and precisely associate a DM to the source environment. 

One way of finding this optimal DM is to repeat measurements on a grid of DMs and choose the DM that minimizes deviations from the sub-burst slope relation, as was done in \citet{Chamma2021}, \citet{Chamma2023}, and \citet{Brown2024}. The main disadvantage of this approach is the time-consuming nature of repeating measurements over many DMs.

A more direct quantitative approach was developed and applied by \citet{Jahns2023} for bursts from FRB 20121102A, by assuming that small burst-to-burst fluctuations in DM are due to drifting behaviour (as opposed to changes in ion column density) and that the sub-burst slope relation extends to very sharp short duration bursts. As our results for ultra-FRBs strongly bear this last assumption out (Figure \ref{fig:nudtdnu}), we will apply this scheme to the measurements presented here. 

Following \citet{Jahns2023}, a fit of the form $\text{d}t/\text{d}\nu \equiv d_{t}(\sigma_{t})=-b\sigma_{t}+c$ is found for the group of burst measurements from a source that have been obtained at the same DM ($\text{DM}_\text{applied}$). Note that these fits are similar to the form of fits tabulated in Table \ref{tab:fits}, except without the multiplication by the frequency $\nu$. Then, interpreting the fit parameter $c$ as `residual' intra-burst drift, the resulting deviation from the real DM, $\delta\text{DM}$, is calculated according to 
\begin{equation}
\delta\text{DM}=-\frac{1}{2}\frac{\nu^{3}}{a_\text{DM}}d_{t},
\end{equation}
where $a_\text{DM}$ is the dispersion constant, $c$ is substituted for $d_{t}$, and we take $\nu$ as the mean burst frequency for the dataset considered. Note that this $\delta$DM can also be found on a burst by burst basis, and used to remove residual drift from each measurement. We performed this analysis, and found similar relationships to the ones already reported, albeit with larger uncertainties on the slopes as bursts were more vertical. In the grouped scheme, we expect $b$ to remain roughly the same since it corresponds (up to a factor of $\nu$) to the $a$ parameter in the sub-burst slope relation, and we expect $c$ to get smaller as the data gets closer to the `real' DM. The `real' DM is then found through
\begin{equation}
\text{DM}_{\text{real}}=\text{DM}_{\text{applied}}+\delta\text{DM}.
\end{equation}
We will repeat this process iteratively, and the obtained $\text{DM}_{\text{real}}$ is used as the applied DM in the following stage. 

We apply this process for the sources FRB 20121102A and FRB 20220912A separately. Note that we exclude the many bursts from \citet{Hewitt2022} for FRB 20121102A due to their different applied starting DM and burst B1 from \citet{Hewitt2023} for FRB20220912A due to the presence of significant blending. The starting DM used is the DM of the shortest (or nearly the shortest) duration burst from that source. In the case of FRB 20121102A that burst is the microburst B30 of \citet{Snelders2023}, and burst B1 of \citet{Hewitt2023} for FRB 20220912A. The results of this are shown in Table \ref{tab:dm}.

We see that the value of $c$ can increase after each round. Though the value for $c$ from FRB 20121102A decreases after the second round, both sets of measurements end up with a larger value of $c$ than they started with. This behaviour is contrary to expectations and may indicate a divergent solution. The value of $b$ for each dataset is consistent with the sub-burst slope relation fits (Table \ref{tab:fits}) once scaled by the mean frequency for that dataset. The value of $\text{DM}_{\text{real}}$ decreases in each round from the initial DM. A possible interpretation of these results is that the initial DM applied is sufficiently optimized that this process does not give much improvement, though for FRB20121102A our DM value of 560.105 pc/cm$^3$ is  different from the value obtained in \citet{Jahns2023}, which was 562.41 pc/cm$^3$. 

To check if we may be in a local minimum, we repeated this analysis at a different starting DM and applied two steps of DM optimization for the measurements from FRB 20121102A. We started at the DM of 562.41 pc/cm$^3$ found by \citet{Jahns2023} and repeated our measurements. The results are in Table \ref{tab:dm}, denoted by steps 1b, 2b, and 3b. We note that the value of residual drift $c$ for our measurements at this DM was $2.0\times10^{-4}$ ms/MHz, larger than for the measurements at 560.105 pc/cm$^3$. We found $\delta\text{DM}$s that again decrease $\text{DM}_\text{real}$ every step, this time by larger steps. The value of $c$ ends up increasing throughout this process, again potentially indicating divergence. Our interpretation of these results is that the DM optimization process is best attempted with large samples sizes of measurements, perhaps hundreds or thousands of bursts. Otherwise, numerical instabilities as seen here can likely occur. The other possibility is that there are sources of intra-burst drift not characterized by the sub-burst slope relation.

Despite the time consuming efforts of repeating and confirming measurements at different DMs, using the sub-burst slope relation with hundreds or thousands of bursts emitted within a similar epoch to associate a precise DM to a source is potentially an important technique for probing the source environment of a repeating FRB.

\section{Figures with Adjusted Methodologies}\label{subsec:figs}

\begin{figure}
    \centering
    \includegraphics[width=\columnwidth]{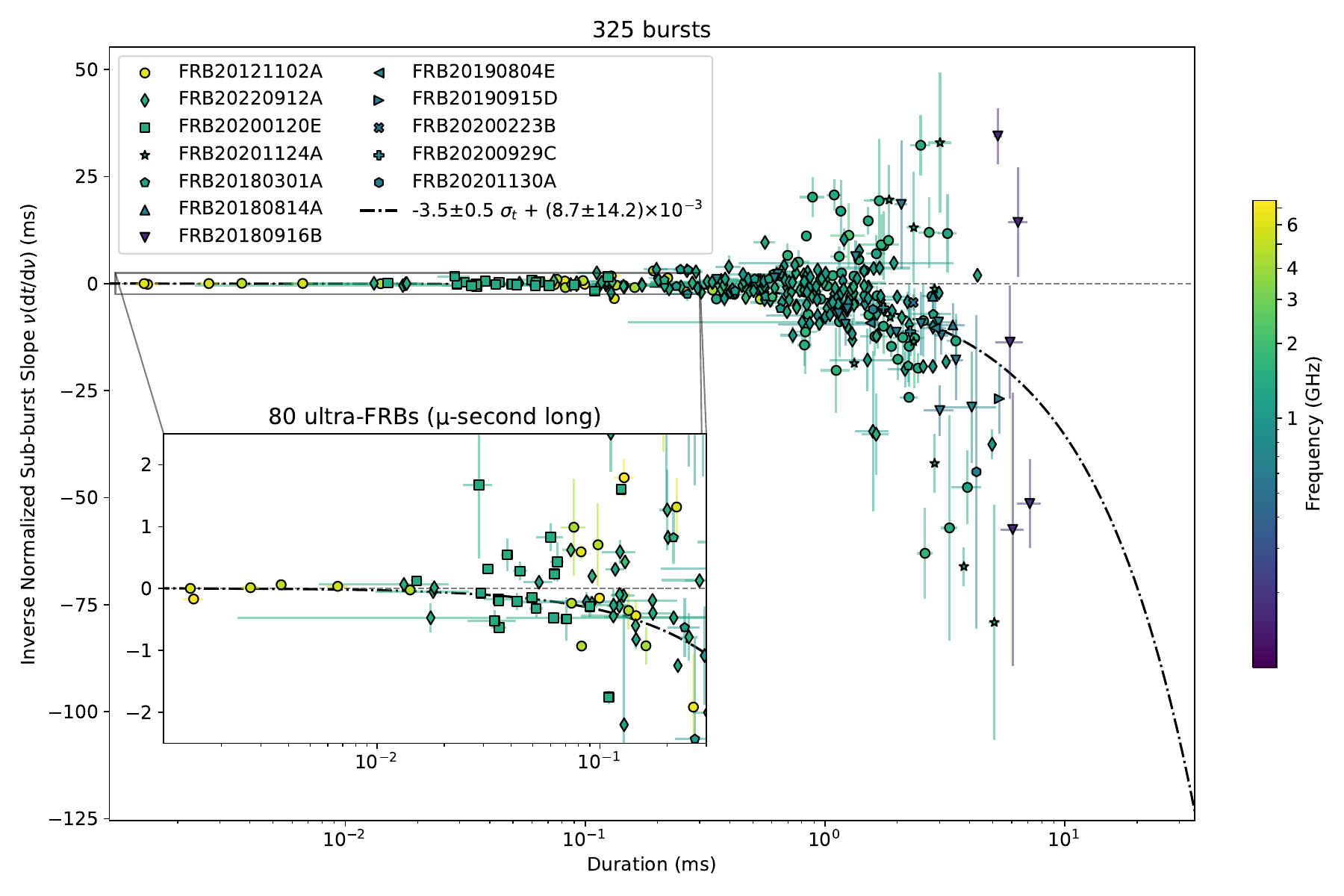}
    \caption{Figure \ref{fig:nudtdnu} of the main text but with bursts measured at their individually determined DMs. See Section \ref{subsec:indivdms} for details.}
    \label{fig:nudtdnu_burstdm}
\end{figure}

\begin{figure}
    \centering
    \includegraphics[width=\columnwidth]{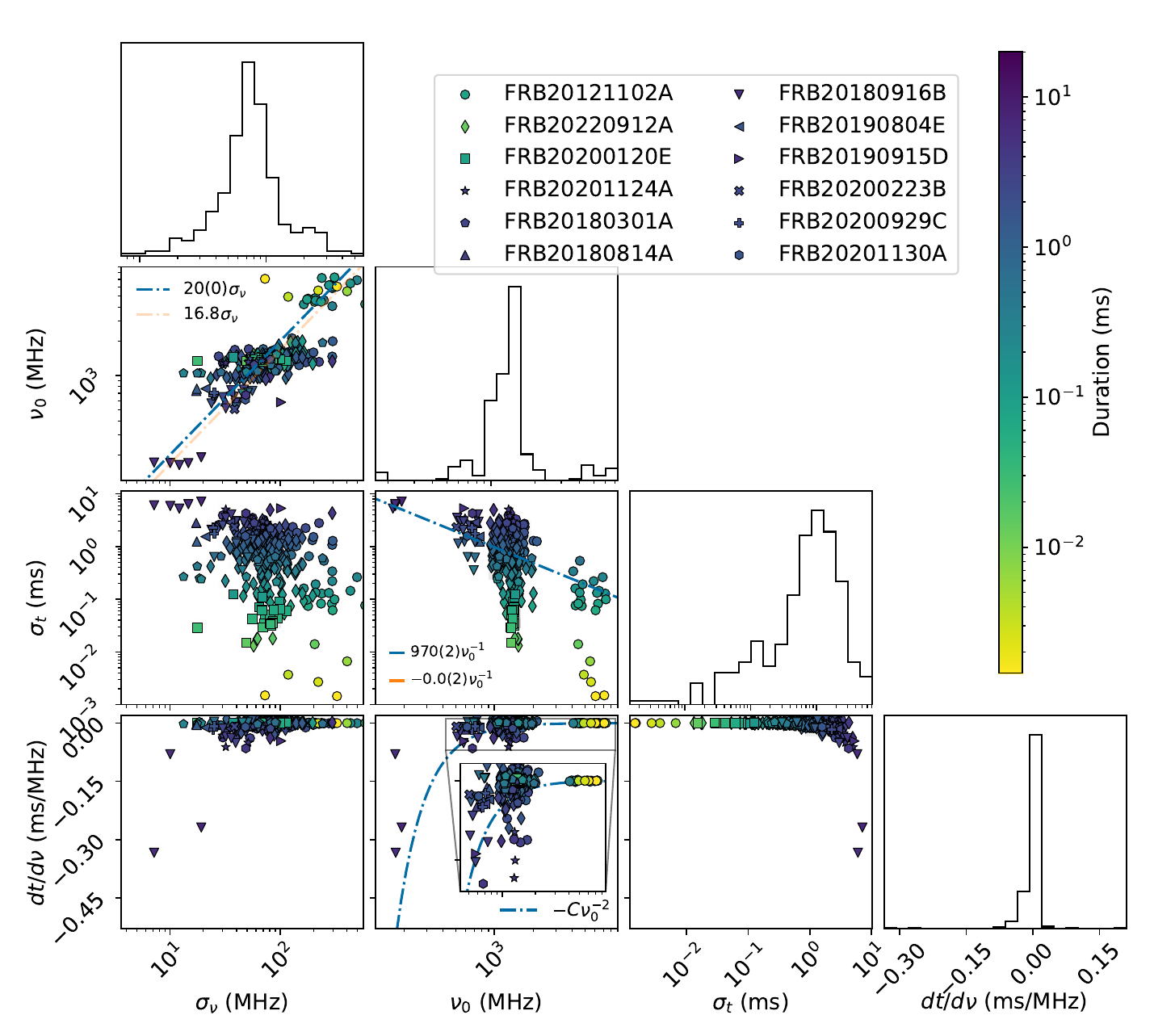}
    \caption{Figure \ref{fig:corner} of the main text but with bursts measured at their individually determined DMs. See Section \ref{subsec:indivdms} for details.}
    \label{fig:corner_burstdm}
\end{figure}

\begin{figure}
    \centering
    \includegraphics[width=\columnwidth]{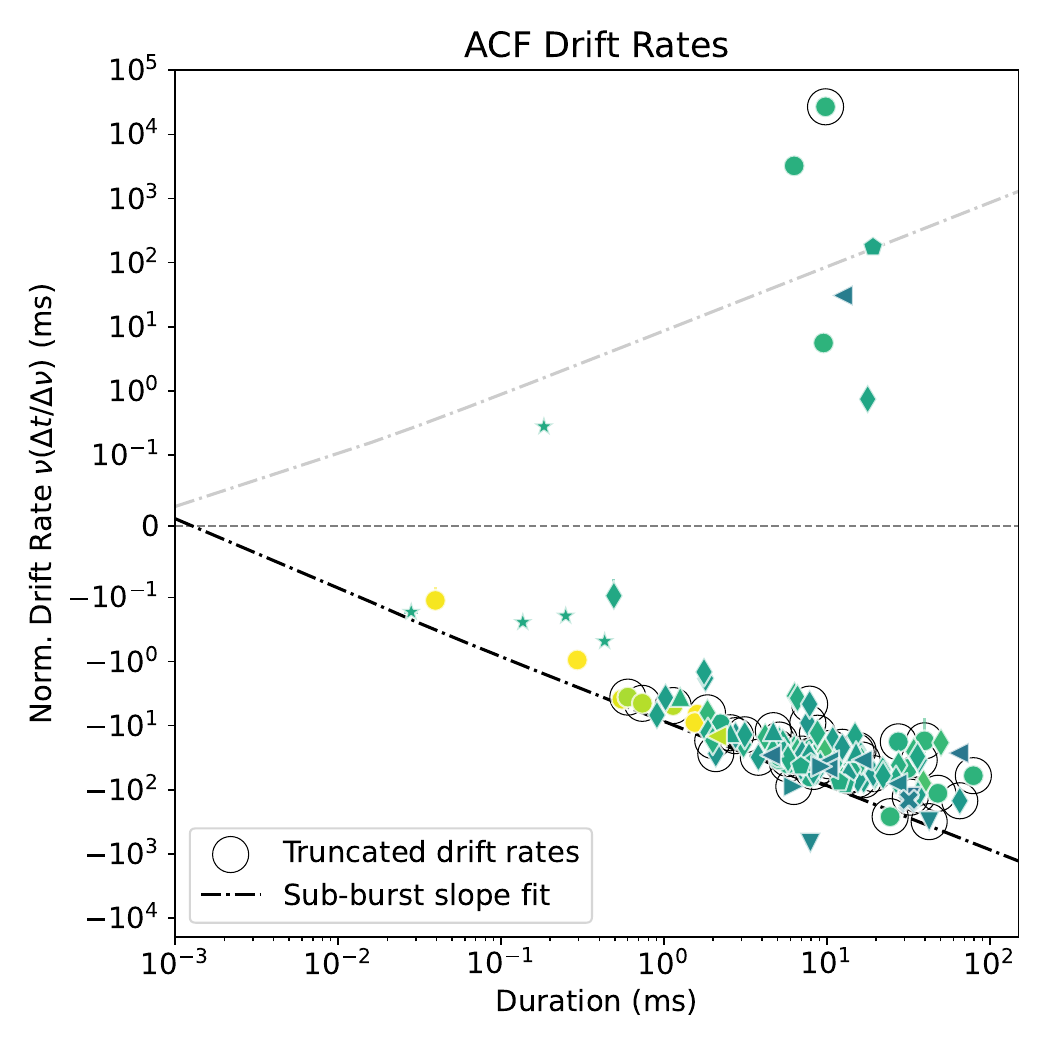}
    \caption{Figure \ref{fig:drift} of the main text but with multi-component drift rates measured using the ACF method. See Section \ref{sec:drifts} for details.}
    \label{fig:driftacf}
\end{figure}

We include here Figures \ref{fig:nudtdnu_burstdm}, \ref{fig:corner_burstdm}, and \ref{fig:driftacf}. These figures are copies of Figures \ref{fig:nudtdnu}, \ref{fig:corner}, and \ref{fig:drift} but with adjustments in the methodology used. Figures \ref{fig:nudtdnu_burstdm} and \ref{fig:corner_burstdm} display measurements from each burst's individually determined DM, and are discussed in detail in Section \ref{subsec:indivdms}. Figure \ref{fig:driftacf} presents multi-component drift rates measured using the ACF method to contrast with the arrival times drift rate measurements. This figure is discussed in detail in Section \ref{sec:drifts}.

\bsp	% typesetting comment
\label{lastpage}
\end{document}